\documentclass[aps,pra,reprint,twocolumn,superscriptaddress,nofootinbib,amsmath,amssymb,longbibliography,preprintnumbers]{revtex4-2}

\usepackage{xcolor}
\usepackage{amsfonts}
\usepackage{mathtools}
\usepackage{graphicx}
\graphicspath{{./}}
\usepackage{booktabs}
\usepackage{empheq}
\usepackage{enumitem}
\usepackage{float}
\makeatletter
\let\newfloat\newfloat@ltx
\let\float@end\float@end@ltx
\makeatother
\usepackage{algorithm}
\usepackage{algpseudocode}
\usepackage{hyperref}
\hypersetup{colorlinks=true,linkcolor=blue!50!black,citecolor=blue!50!black,urlcolor=blue!50!black}
\usepackage[capitalize,nameinlink]{cleveref}
\crefname{section}{Sec.}{Secs.}      \Crefname{section}{Sec.}{Secs.}
\crefname{appendix}{Appendix}{Appendices} \Crefname{appendix}{Appendix}{Appendices}
\crefname{equation}{Eq.}{Eqs.}       \Crefname{equation}{Eq.}{Eqs.}
\crefname{table}{Table}{Tables}      \Crefname{table}{Table}{Tables}
\crefname{figure}{Fig.}{Figs.}       \Crefname{figure}{Fig.}{Figs.}
\crefname{algorithm}{Algorithm}{Algorithms} \Crefname{algorithm}{Algorithm}{Algorithms}
\newcommand{\GF}{\mathbb{F}_2}
\newcommand{\mwpm}{\textsc{mwpm}}
\newcommand{\bposd}{\textsc{bp-osd}}
\newcommand{\obf}{\textsc{obf}}
\newcommand{\zk}{Z-\textsc{obf}}
\newcommand{\ck}{\textsc{ce}}
\newcommand{\zh}{\zk{}+\ck{}}
\newcommand{\iat}{\hat\tau}

\DeclareMathOperator{\im}{im}

\begin{document}

\title{Fast and accurate rare-event simulation by Markov chain Monte Carlo for general quantum codes}

\author{Junichi Haruna}
\affiliation{Graduate School of Informatics, Kyoto University, Sakyo-ku, Kyoto 606-8501, Japan}
\affiliation{Center for Quantum Information and Quantum Biology, University of Osaka, Toyonaka, Osaka 560-0043, Japan}
\email{j.haruna1111@gmail.com}

\author{Kotaro Murakami}
\affiliation{Department of Physics, Institute of Science Tokyo, 2-12-1 Ookayama, Meguro-ku, Tokyo 152-8551, Japan}
\affiliation{RIKEN Center for Interdisciplinary Theoretical and Mathematical Sciences (iTHEMS), RIKEN, Wako, Saitama 351-0198, Japan}
\email{kotaro.murakami@yukawa.kyoto-u.ac.jp}

\date{\today}
\preprint{RIKEN-iTHEMS-Report-26}

\begin{abstract}
The splitting method estimates logical error rates of quantum error-correcting codes deep in the rare-event regime, where direct Monte Carlo sampling is prohibitively expensive.
It relies on a Markov chain Monte Carlo (MCMC) sampler of faulty configurations, usually driven by one-bit-flip proposals.
However, one-bit flips have been reported to be trapped on some codes, which biases the estimate.
To address this issue, we propose a new kernel that combines locally balanced one-bit-flip proposals with a cluster-exchange move.
We benchmark both kernels and direct Monte Carlo on twelve codes under code-capacity bit-flip noise in two ways, by comparing the mean fault weight with an MCMC-independent reference and by measuring the computation time to reach a fixed accuracy.
In the first, our kernel agrees with the reference on every code, including those where one-bit flips are trapped.
In the second, it is faster than one-bit flips wherever they are trapped, by orders of magnitude on some codes.
Furthermore, at the lowest error rate, it is faster than direct Monte Carlo on every code, by more than four orders of magnitude on some.
\end{abstract}

\maketitle

\section{Introduction}
\label{sec:intro}

Scaling quantum computation to practically relevant sizes is obstructed by the fragility of quantum information under noise.
Quantum error correction (QEC) overcomes this obstacle by encoding logical information redundantly into a larger physical system so that errors can be detected and corrected~\cite{gottesman1997stabilizer,fowler2012surface}.
Many code families now compete for this role, including topological codes~\cite{kitaev2003fault,dennis2002topological,fowler2012surface}, concatenated codes~\cite{steane1996multiple,aliferis2006accuracy}, and quantum low-density parity-check (qLDPC) codes~\cite{panteleev2021degenerate,bravyi2024highthreshold}.
How well a given code and decoder protect the logical information is measured by one number, the logical error rate $P_L$, which is the probability that error correction ends in a logical failure.
Practical fault-tolerant algorithms demand values as small as $P_L \sim 10^{-12}$~\cite{beverland2025failfast}.

The standard way to estimate the logical error rate $P_L$ is direct Monte Carlo, which samples error configurations from the noise model, decodes each one, and counts the fraction of logical failures.
This estimate is unbiased by construction, but it needs $1/(\epsilon^{2}P_L)$ samples for a fixed relative accuracy $\epsilon$.
Failures of a good code are rare events at the physical error rates $p \sim 10^{-3}$--$10^{-4}$ expected of near-term hardware~\cite{fowler2012surface,bravyi2024highthreshold}.
The direct estimate therefore becomes prohibitively expensive for codes of practical size exactly in the regime where its answer matters most.
Estimating the logical error rate there requires a dedicated rare-event method.

The splitting method~\cite{bravyi2013simulation,mayer2025rare} is a rare-event estimator developed for this problem.
It writes the logical error rate $P_L$ as its value at a high error rate multiplied by a ladder of ratios between progressively lower error rates.
Each ratio is an expectation over the conditional distribution of faulty configurations at a nearby error rate, and it is estimated from Markov chain Monte Carlo (MCMC) samples of that distribution.
Bravyi and Vargo introduced the method for the surface code~\cite{bravyi2013simulation}.
It has since been scaled to circuit-level noise and has reached simulated failure rates below $10^{-20}$~\cite{mayer2025rare}.

Yet the entire construction rests on its Metropolis--Hastings sampler~\cite{metropolis1953,hastings1970}.
The cost of the ladder is set by how fast the chains \emph{mix}, that is, how many steps a chain needs before it forgets its initial state and its samples become effectively independent.
This number sets the chain length required for a given accuracy, whereas the correctness of the ladder is set by whether the chains cover the faulty configurations at all.

Most previous implementations have driven the chains with proposals that flip a single bit~\cite{bravyi2013simulation,mayer2025rare,bravyi2024highthreshold,beverland2025failfast}.
Bravyi and Vargo already conjectured that such local moves mix slowly when low-weight faulty configurations in different logical sectors can only be connected through high-weight detours~\cite{bravyi2013simulation}.
Beverland et al.\ recently reported this trapping for the asymmetric toric code, where their chains initialized in one logical sector fail to reach the others~\cite{beverland2025failfast}.
To mitigate it, they seed the chains with faulty configurations drawn by direct Monte Carlo at the top of the ladder and pass the final state of each level to the next.
Both keep the one-bit-flip proposal, and they leave proposals beyond one-bit flips to future work.

The one-bit flip is only the simplest choice of proposal.
Other applications of MCMC use richer kernels to accelerate slowly mixing chains, from cluster updates in statistical physics~\cite{newman1999monte} to informed proposals in computational statistics~\cite{zanella2020informed,livingstone2022barker}.
This suggests that a better kernel can speed up the splitting method as well.
Mullan et al.\ took a first step in this direction very recently~\cite{mullan2026improved}.
They resample whole subregions of the error pattern at once and report faster convergence than one-bit flips on surface and concatenated Bacon--Shor codes.
However, it has not been established whether any of these kernels samples the faulty configurations of general quantum codes efficiently and without trapping, in particular on the code families where the trapping is observed.

In this work, we propose a kernel for the splitting method that is fast and that, on every code we test, agrees with a reference value computed without MCMC, including the code families on which one-bit flips are trapped.
The kernel combines two moves.
The first is a locally balanced one-bit-flip proposal in the sense of Zanella~\cite{zanella2020informed}, which weights the candidate flips toward those likely to be accepted instead of flipping a uniformly random bit, so that the random walk within a logical sector wastes fewer proposals.
The second is a cluster-exchange move, which moves a few errors to other qubits at fixed weight.
Because this move preserves the weight, it can carry the chain between logical sectors without ever climbing the high-weight barrier that obstructs one-bit flips.
Our focus is the code-capacity setting under independent bit-flip noise, where a reference value computed without MCMC is available.
There, the logical error rate and the mean fault weight can be computed by enumerating or uniformly sampling the faulty configurations at each error weight and summing over the weights (\Cref{sec:setup-stratified}).

We benchmark this kernel at physical error rates from $p=10^{-1}$ down to $10^{-5}$ on codes spanning surface~\cite{dennis2002topological,fowler2012surface} and toric codes~\cite{kitaev2003fault}, asymmetric toric codes~\cite{kitaev2003fault,dennis2002topological}, the concatenated Steane code~\cite{steane1996multiple,aliferis2006accuracy}, a hypergraph product (HGP) code~\cite{tillich2014quantum} built from classical Hamming codes, HGP(Hamming), and bivariate bicycle (BB) codes~\cite{bravyi2024highthreshold}.
We run two benchmarks, one on the accuracy of the mean fault weight and one on the time to accuracy.
On twelve codes, the first runs each kernel at a fixed budget from up to two different initial states and compares its estimate with an MCMC-independent reference.
It thus tests a necessary condition for unbiased sampling of the fault manifold, namely agreement of the mean fault weight with the reference.
The second measures the computation time each kernel needs to estimate the mean fault weight to $2\%$ relative accuracy, judged against the same reference, on the same twelve codes.
A chain whose bias exceeds $2\%$ of the reference never reaches the target, so the second benchmark catches such a kernel as well as a slow one.

Our results are threefold.
First, in the rare-event regime at $p=10^{-5}$, our kernel estimates the mean fault weight faster than direct Monte Carlo on all twelve benchmark codes, with advantages exceeding four orders of magnitude on some codes.
Under stated assumptions, we estimate in \Cref{sec:plcost} that a similar advantage carries over to estimating the logical error rate.
There, the splitting method with our kernel is estimated to reach $2\%$ relative accuracy on the logical error rate at $p=10^{-5}$ more cheaply than direct Monte Carlo on all twelve codes, by a factor of about $10$ or more.
Second, on codes whose logical operators all have nearly the same minimum weight (in a given basis), both the one-bit-flip baseline and our kernel reproduce the reference well with sufficient accuracy, and ours costs only a constant factor more per step.
These codes are the surface and toric codes, the concatenated Steane code, and the Hamming hypergraph product code.
Third, on codes whose logical operators differ widely in weight, the asymmetric toric and BB codes, the baseline is trapped.
Its estimate of the mean fault weight is biased from at least one initial state on all of these codes except the $[[72,12,6]]$ BB code, and our kernel removes the bias.
At the error rates $p=10^{-3}$ and $10^{-5}$, our kernel reaches the accuracy target at least $500$ times sooner than the baseline on the asymmetric toric codes, and $21$ and at least $10^{3}$ times sooner on the $[[72,12,6]]$ BB code, respectively.

\begin{figure}[t]
  \centering
  \includegraphics[width=\linewidth]{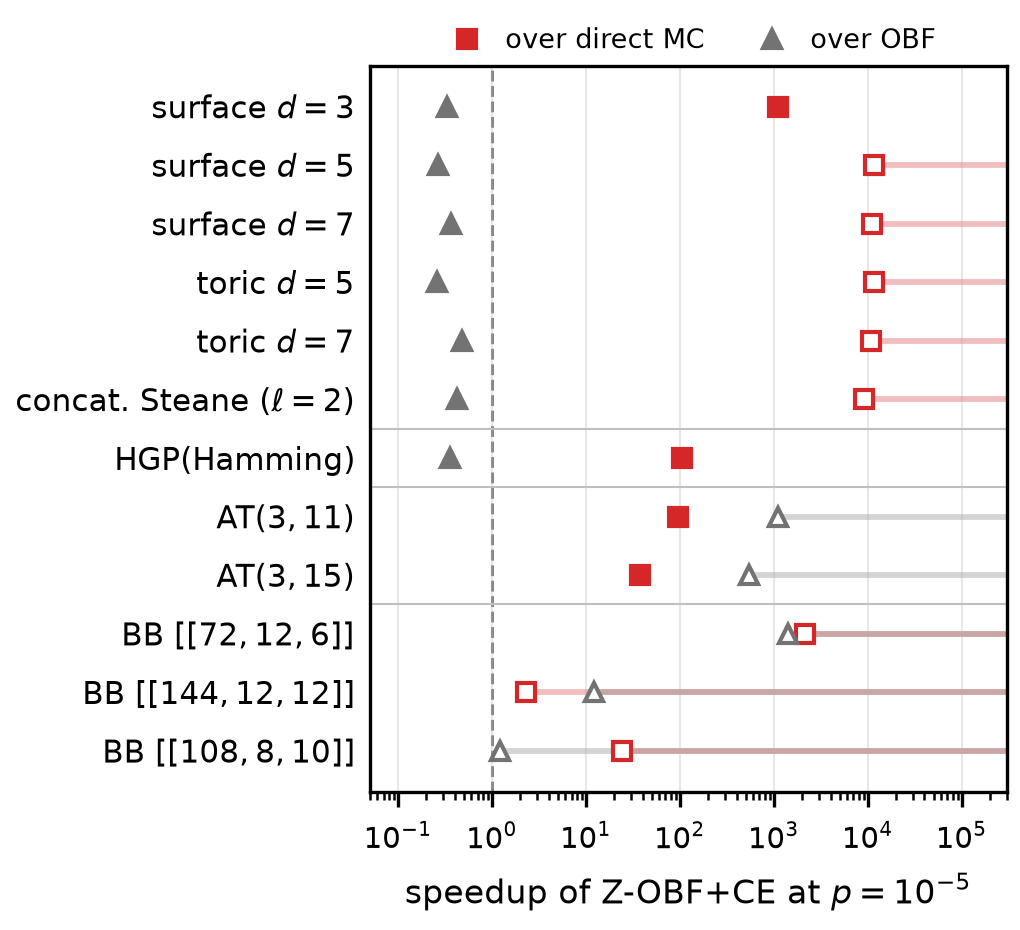}
  \caption{Speedup of our kernel, Zanella one-bit flips with cluster exchange (\zh{}), at $p=10^{-5}$ on the twelve codes, the time to accuracy $T_{2\%}$ (\Cref{sec:bench2}) of direct Monte Carlo (squares) or of the one-bit-flip kernel \obf{} (triangles) divided by that of \zh{}, from \Cref{tab:dmcspeedup,tab:obfspeedup}.
  An open marker with a line to the right is a lower bound, where the slower method did not reach the target within its budget.
  The standard errors defined in \Cref{sec:bench2} are smaller than the markers.
  The dashed line marks equal time.}
  \label{fig:intro}
\end{figure}

Our results have one message.
Rare-event splitting can estimate the logical error rate at error rates far below the reach of direct sampling, provided that its kernel explores the faulty configurations of the code both without bias and quickly.
With such a kernel, the method is at once fast and accurate, and our kernel meets both requirements on every code we test.
\Cref{fig:intro} shows the speedup of our kernel on all twelve codes at the lowest physical error rate.

The remainder of this paper is organized as follows.
In \Cref{sec:prelim}, we collect the background on Calderbank--Shor--Steane (CSS) codes, the logical error rate and its direct Monte Carlo estimator, and the splitting method with its one-bit-flip baseline.
In \Cref{sec:kernel}, we construct the kernel.
In \Cref{sec:setup}, we fix the numerical setup, from the code list to the estimation conventions.
In \Cref{sec:results}, we show against an MCMC-independent reference that the baseline is trapped and our kernel shows no comparable trapping bias within the tested budgets, and we measure the computation time of both kernels on every code.
In \Cref{sec:discussion}, we turn the results into a recipe for verifying a new code, evaluate the cost of estimating the logical error rate from the numerical results, and state the conditions under which the method extends to other codes, decoders, and noise models.
In \Cref{sec:conclusion}, we conclude with an outlook.

\section{Preliminaries}
\label{sec:prelim}

\subsection{Calderbank--Shor--Steane (CSS) codes}
\label{sec:css}

The stabilizer formalism~\cite{gottesman1997stabilizer} encodes $k$ logical qubits into $n$ physical qubits via a commuting Pauli stabilizer group $\mathcal{S}$, and the code space is the joint $+1$ eigenspace of all stabilizers.
The code distance $d$ is the minimum weight of any Pauli operator that commutes with the stabilizer group $\mathcal{S}$ but is not itself a stabilizer up to phase, that is, of any nontrivial logical operator (defined below).

A Calderbank--Shor--Steane (CSS) code~\cite{calderbank1996good,steane1996multiple} is defined by two binary parity-check matrices $H_X \in \GF^{r_X\times n}$ and $H_Z \in \GF^{r_Z\times n}$, with $r_X$ and $r_Z$ rows, respectively, and one column per physical qubit, satisfying $H_Z H_X^\top = 0 \pmod{2}$.
The $r_X$ stabilizer generators of $X$ type are $X((H_X)_i)$, one per row of the parity-check matrix $H_X$, and the $r_Z$ generators of $Z$ type are $Z((H_Z)_i)$, one per row of $H_Z$, where
\begin{align}
  X(v) \coloneqq \prod_{i=1}^n X_i^{v_i}, \qquad
  Z(v) \coloneqq \prod_{i=1}^n Z_i^{v_i}
\end{align}
are the Pauli operators corresponding to a binary vector $v \in \GF^n$.
An $X$-type error $e \in \GF^n$ produces the $Z$-syndrome $H_Z e \pmod{2}$.
Similarly, $Z$-type errors produce $X$-syndromes via the parity-check matrix $H_X$.
In CSS codes, the two syndromes decouple, so under independent $X$ and $Z$ noise, the decoding problem separates exactly.

Logical Pauli operators are Pauli operators that commute with all stabilizers but are not themselves stabilizers.
We fix two logical matrices $L_X,L_Z\in\GF^{k\times n}$ whose row cosets form bases of the quotient spaces $\ker H_Z/\im H_X^\top$ and $\ker H_X/\im H_Z^\top$, respectively, and write the logical Pauli operators as $\overline{X}_i=X((L_X)_i)$ and $\overline{Z}_i=Z((L_Z)_i)$.

\subsection{Fault manifold and direct Monte Carlo}
\label{sec:decoding}

A decoder $\mathcal{D}$ maps the measured syndrome $H_Z e$ to a correction $\mathcal{D}(H_Z e)$ with the same syndrome.
For an $X$-type error $e$, the residual error $e+\mathcal{D}(H_Z e)$ therefore lies in the kernel of $H_Z$, and decoding fails when that residual error represents a nontrivial logical operator.
The corresponding \emph{fault manifold} is defined as
\begin{align}
  \mathcal{F} \coloneqq \{e \in \GF^n : e + \mathcal{D}(H_Z e) \in \ker H_Z \setminus \im H_X^\top\}.
\end{align}
The residual error of every faulty configuration represents one of the $2^k-1$ nontrivial classes of the quotient space $\ker H_Z/\im H_X^\top$.
Faulty configurations whose residual errors belong to the same equivalence class form a \emph{logical sector}.
The fault manifold is thus the disjoint union of $2^k-1$ logical sectors.
We note that the fault manifold depends on the decoder and its parameters as well as on the code.
In particular, it can change with the decoder settings, such as the physical error rate that belief propagation with ordered statistics decoding (\bposd{})~\cite{panteleev2021degenerate} takes as an input (\Cref{sec:scope}) or the update rule of its belief propagation~\cite{fregona2026lowweight}.

Under a given noise model, each error configuration $e$ occurs with some probability $P(e)$.
The \emph{logical error rate} $P_L$, which measures the performance of the code--decoder pair, is defined as the total probability of the faulty configurations,
\begin{align}
  P_L &\coloneqq \Pr(\mathcal{F}) = \sum_{e \in \mathcal{F}} P(e).
\end{align}

Besides the logical error rate $P_L$, we use the mean fault weight to test how well a sampler explores the fault manifold $\mathcal{F}$.
Let $w(e)\coloneqq|e|$ be the Hamming weight of a configuration $e$, the number of flipped qubits, and let $w$ denote the same quantity regarded as an observable on the fault manifold.
The mean fault weight is defined as the conditional expectation of the weight over the fault manifold,
\begin{align}
  \mathbb{E}_{\mathcal{F}}[w] \coloneqq \frac{1}{P_L}\sum_{e \in \mathcal{F}} w(e)\,P(e),
  \label{eq:meanw}
\end{align}
and the subscript $\mathcal{F}$ marks this conditioning throughout.
A kernel that misestimates this simple observable cannot be trusted to sample the fault manifold, and we therefore use this observable to test every kernel.

The most elementary estimator of the logical error rate $P_L$ is \emph{direct (rejection) Monte Carlo}.
We draw a configuration $e$ from the noise distribution $P(e)$, calculate the residual error $e + \mathcal{D}(H_Z e)$, and report the fraction $\hat P_L$ of faulty draws among all draws.
Because the draws are independent and identically distributed (i.i.d.)\ from the noise distribution, the estimator $\hat P_L$ is unbiased with an easily computed variance, and the same faulty draws yield the mean fault weight $\mathbb{E}_{\mathcal{F}}[w]$ as their average weight.

Its cost is governed by the rarity of faulty draws.
Achieving relative error $\epsilon$ requires $N_{\mathrm{trial}}\sim1/(\epsilon^2P_L)$ trials, each a decoder call, so direct Monte Carlo becomes prohibitive in the rare-event regime.

\subsection{Rare-event estimation by the splitting method}
\label{sec:splitting}

The splitting method~\cite{bravyi2013simulation,mayer2025rare,beverland2025failfast} circumvents the $O(1/P_L)$ cost of direct Monte Carlo by decomposing the logical error rate $P_L$ into a product of ratios, each of which we estimate by the Markov chain Monte Carlo method.

Under independent bit-flip noise, each qubit flips with probability $p$, giving
\begin{align}
  \pi_{p}(e) = p^{w(e)}(1-p)^{n-w(e)}.
\end{align}
Let $p_0 > p_1 > \cdots > p_J = p$ be a decreasing sequence of error rates chosen so that the logical error rate $P_L(p_0)$ is accessible by direct Monte Carlo.
The logical error rate at level $\ell$ is then written as
\begin{align}
  P_L(p_\ell) = P_L(p_0) \prod_{j=0}^{\ell-1} R_j,
  \qquad
  R_j \coloneqq \frac{P_L(p_{j+1})}{P_L(p_j)}.
\end{align}
Thus, only the accessible logical error rate $P_L(p_0)$ and the adjacent ratios $R_j$ must be estimated.
We obtain each ratio from the conditional fault distribution
\begin{align}
  \pi_{p_\ell\mid\mathcal{F}}(e) \coloneqq \frac{\pi_{p_\ell}(e)}{P_L(p_\ell)}, \qquad e\in\mathcal{F},
  \label{eq:condfault}
\end{align}
with zero probability outside the fault manifold $\mathcal{F}$.
Bravyi and Vargo~\cite{bravyi2013simulation} proposed estimating this ratio with Bennett's acceptance ratio (BAR) method~\cite{bennett1976efficient}.
For any constant $C > 0$, the ratio admits the representation
\begin{align}
  R_\ell = C\,
  \frac{
    \mathbb{E}_{\pi_{p_\ell\mid\mathcal{F}}}\left[h\left(C\,
      \dfrac{\pi_{p_\ell}(\cdot)}{\pi_{p_{\ell+1}}(\cdot)}\right)\right]}
    {
    \mathbb{E}_{\pi_{p_{\ell+1}\mid\mathcal{F}}}\left[h\left(\dfrac{1}{C}\,
      \dfrac{\pi_{p_{\ell+1}}(\cdot)}{\pi_{p_\ell}(\cdot)}\right)\right]},
  \label{eq:bar}
\end{align}
where the function $h$ satisfies $h(1/t)=t h(t)$, and we use the variance-optimal Fermi function $h(t)=1/(1+t)$~\cite{bennett1976efficient}.
Setting the constant $C$ to the ratio $R_\ell$ yields a self-consistent equation whose expectations we estimate with MCMC samples from the two adjacent conditional fault distributions.

The Metropolis--Hastings method~\cite{metropolis1953,hastings1970} samples these distributions by proposing a configuration $e'$ from a proposal distribution $q(\cdot\mid e)$ and accepting it with probability
\begin{align}
  A(e,e') \coloneqq \min\!\left(1,\;
    \frac{\pi_{p}(e')\,q(e\mid e')}{\pi_p(e)\,q(e'\mid e)}\right).
  \label{eq:mh}
\end{align}
For the conditional fault distribution of \Cref{eq:condfault}, the unknown normalizer $P_L(p_\ell)$ cancels from the acceptance probability, and proposals outside the fault manifold $\mathcal{F}$ are rejected.
We call a proposal distribution $q$ together with this acceptance rule a \emph{kernel}.
Every kernel compared here uses the same acceptance rule, so the kernels differ only in the proposal distribution $q$.
\Cref{alg:mh} gives the full procedure.

Two properties of a kernel decide whether, and how fast, its averages over the steps approach the expectations under the conditional fault distribution.
The first is ergodicity.
A kernel is \emph{ergodic} on the fault manifold if the chain can reach every faulty configuration from every other one in a finite number of accepted moves.
For an ergodic kernel, the Metropolis--Hastings acceptance rule guarantees that the averages over the steps converge to these expectations as the chain grows.
Ergodicity depends on the proposal distribution $q$ and on the fault manifold, so it has to be established for each combination of code, decoder, and kernel.
A nonergodic kernel converges instead to an average over the reachable part of the fault manifold, so its estimate could remain biased however long the chain runs.

The second is autocorrelation.
Even for an ergodic kernel, successive states of the chain are correlated, since each state is generated from the previous one through the proposal distribution $q$.
The \emph{autocorrelation time} $\tau$, defined in \Cref{sec:bench1}, is the number of steps over which this correlation decays, so an average over $N$ steps has the variance of $N/\tau$ independent draws.
Autocorrelation therefore sets the cost of a given accuracy, whereas ergodicity sets whether that accuracy is attainable at all.

\begin{algorithm}[t]
\caption{Metropolis--Hastings sampling on the fault manifold $\mathcal{F}$ at level $p_\ell$}
\label{alg:mh}
\begin{algorithmic}[1]
\Require initial faulty state $e_0 \in \mathcal{F}$, proposal $q$, number of steps $N$
\Ensure samples $\{e_t\}_{t=1}^{N} \subset \mathcal{F}$ targeting $\pi_{p_\ell\mid\mathcal{F}}$
\State $e \gets e_0$
\For{$t = 1,\dots,N$}
  \State propose $e' \sim q(\cdot\mid e)$
  \State $a \gets \min\!\left(1,\ \dfrac{\pi_{p_\ell}(e')\,q(e\mid e')}{\pi_{p_\ell}(e)\,q(e'\mid e)}\right)$
  \State draw $u \sim \mathrm{Uniform}(0,1)$
  \If{$u \le a$}
    \State decode $e'$ to test $e' \in \mathcal{F}$
    \If{$e' \in \mathcal{F}$}
      \State $e \gets e'$
    \EndIf
  \EndIf
  \State $e_t \gets e$
\EndFor
\State \Return $\{e_t\}_{t=1}^{N}$
\end{algorithmic}
\end{algorithm}

\subsection{The one-bit-flip baseline}
\label{sec:baseline}

The one-bit-flip kernel, the baseline of this paper, chooses one of the $n$ bits uniformly and flips it, so that its proposal distribution is given by
\begin{empheq}[left={q(e'\mid e) =\empheqlbrace}]{align}
   &\frac{1}{n}, \qquad (e' = e + u_i,\ i=1,\dots,n) \\
   &0, \qquad (\text{otherwise}),
\end{empheq}
where $u_i$ is the unit vector with a $1$ in the $i$th position and $0$ elsewhere.
Here and below, we drop the level index and write $p$ for the error rate $p_\ell$ and $\pi_p$ for the bit-flip distribution $\pi_{p_\ell}$ for notational simplicity.
Since the proposal is symmetric, the acceptance probability of \Cref{eq:mh} reduces to
\begin{align}
  A(e,e') &= \min\!\left(1,\ \frac{\pi_{p}(e')}{\pi_{p}(e)}\right) \\
  &= \min \left(1,\left(\dfrac{p}{1-p} \right)^{\Delta w}\right),
\end{align}
with the weight change $\Delta w = w(e')-w(e)$.
Each proposal costs at most one decoder call, since only those passing the acceptance test are decoded to check whether the proposed configuration $e'$ lies in the fault manifold $\mathcal{F}$ (\Cref{alg:mh}).

This simple kernel has been used in the literature for rare-event simulation in quantum error correction.
However, its applicability and performance on general quantum codes raise two concerns.

The first is ergodicity.
It has been proven for the surface code in two geometries with holes under minimum-weight decoding~\cite{bravyi2013simulation}, but no such argument is known for general quantum codes, to our knowledge.
Without ergodicity, the kernel visits only the part of the fault manifold connected to its start, and its estimates depend on the initial state.

The second is efficiency, even where the kernel is ergodic.
At small error rates, the chain spends most of its steps near the minimum weight of its sector, where almost every proposal increases the weight and is accepted with probability only about $p$.
Only a fraction of order $p$ of all steps is accepted, even within one logical sector.
We expect a more serious problem on a code whose logical weights are broadly spread.
We conjecture that the low-weight configurations of different logical sectors are separated by a high-weight barrier on such a code.
That is, every path of single flips between them passes through configurations of much larger weight.
Crossing a barrier of height $\Delta w$ then takes $\Delta w$ consecutive weight-increasing flips, and the chain spends of order $1/p^{\Delta w}$ steps on one crossing.
If this picture is right, the one-bit-flip kernel is trapped within one logical sector on practical budgets, even though it would cross the barrier given enough time.
In \Cref{sec:results}, we test this expectation against an MCMC-independent reference and find that the trapping does occur on codes with a broad spread of logical weights.

\section{A kernel with locally balanced flips and cluster exchange}
\label{sec:kernel}

The previous section reviewed the splitting method and its one-bit-flip baseline (hereafter \obf{}).
That baseline has two limitations.
Its ergodicity is proven only for the surface code in the two geometries of Ref.~\cite{bravyi2013simulation}, and, as we show in \Cref{sec:results}, it is trapped on codes with a broad spread of logical weights.

To address them, we propose a new kernel defined as the mixture of the following two moves.
One is the locally balanced (Zanella) proposal \zk{}, which favors weight-decreasing flips and is intended to speed up the motion within a sector.
The other is the cluster exchange \ck{}, which can carry the chain between sectors without changing the weight $w$.
At each step, \zh{} proposes a cluster exchange with probability $p_{\mathrm{cluster}}$ and a locally balanced flip with probability $1-p_{\mathrm{cluster}}$.
Each move satisfies detailed balance with respect to the conditional fault distribution on its own, and so does the mixture, since the mixing probability does not depend on the current configuration.
We explain each move in turn in the following subsections.

Note that the ergodicity of the new kernel is not proven either, and we test it empirically against an MCMC-independent reference in \Cref{sec:results}.

\subsection{Locally balanced (Zanella) proposals}
\label{sec:zanella}
The \obf{} kernel selects a bit uniformly at random to flip, which is inefficient when the target distribution is strongly concentrated on low-weight configurations.
The locally balanced proposal of Zanella~\cite{zanella2020informed} remedies this inefficiency by weighting the proposal toward moves that are likely to be accepted.
It proposes a one-bit flip $e\to e+u_i$ with probability proportional to a balancing weight $\lambda_i(e)$, which is a function of the ratio of bit-flip probabilities $\pi_p(e+u_i)/\pi_p(e)$.
In our case, the balancing weight of flip $i$ is given by
\begin{align}
  \lambda_i(e) \coloneqq g\left(\frac{\pi_p(e+ u_i)}{\pi_p(e)}\right),
  \label{eq:zanella_weight}
\end{align}
where $g$ is a positive function satisfying $g(t)=t\,g(1/t)$, called a balancing function.
Zanella's locally balanced proposal then draws a configuration $e'$ from the proposal distribution
\begin{align}
  q(e'\mid e) = \frac{\lambda_i(e)}{\Lambda(e)}
  \label{eq:zanella_proposal}
\end{align}
for $e' = e + u_i$ with $i = 1,\dots,n$, and $q(e'\mid e) = 0$ otherwise, where $\Lambda(e)\coloneqq\sum_j \lambda_j(e)$ is the proposal normalizer.

Among the choices of balancing functions, we use the square root $g(t)=\sqrt{t}$ in this work.
The proposal weights of \Cref{eq:zanella_weight} and the normalizer $\Lambda(e)$ then reduce to
\begin{align}
  \lambda_i(e) &= \sqrt{\frac{\pi_p(e+ u_i)}{\pi_p(e)}} = \left(\frac{p}{1-p}\right)^{\!\Delta w_i / 2},
  \label{eq:zanella_weight_sqrt} \\
  \Lambda(e) &= w(e)\sqrt{\frac{1-p}{p}} + \bigl(n-w(e)\bigr)\sqrt{\frac{p}{1-p}},
  \label{eq:zanella_normalizer}
\end{align}
with the weight change $\Delta w_i = w(e+u_i)-w(e) \in \{-1,+1\}$.
From this relation, we see that the proposal favors weight-decreasing moves in the low-error regime $p\ll 1$.

With the square-root balancing, the ratio of bit-flip probabilities and the ratio of balancing weights cancel each other in the Metropolis--Hastings rule, and the acceptance reduces to the normalizers alone,
\begin{align}
  A(e,e') = \min\!\left(1,\ \frac{\Lambda(e)}{\Lambda(e')}\right).
  \label{eq:zanella_accept}
\end{align}
When the physical error rate $p$ is so small that $p/(1-p) \ll w(e)/(n-w(e))$, the normalizer of \Cref{eq:zanella_normalizer} is dominated by its first term.
The acceptance probability of \Cref{eq:zanella_accept} can then be approximated by
\begin{align}
  A(e,e') \approx \min\!\left(1,\ \frac{w(e)}{w(e')}\right).
  \label{eq:zanella_accept_approx}
\end{align}
Since neighboring configurations differ in weight by $\Delta w = \pm 1$, this acceptance is of order unity even for a weight-increasing move, whereas \obf{} accepts such a move with probability $O(p)$.
The locally balanced proposal is thus expected to accelerate the random walk within a logical sector.

Zanella~\cite{zanella2020informed} shows that, when the number of bits is large, locally balanced proposals are Peskun-optimal among the proposals whose weights are functions of the ratio of bit-flip probabilities, that is, they have the smallest asymptotic variance for every observable.
This guarantee, however, does not cover our case, where the target is restricted to the fault manifold.
We therefore adopt the locally balanced proposal as a heuristic and test its performance numerically.

The cost of the locally balanced proposal per step rises compared to \obf{}, because the proposal reaches the decoder more frequently.
As in \Cref{alg:mh}, a proposal is decoded only after it passes the acceptance test, so each step still costs at most one decoder call.
Under \obf{}, a weight-increasing flip is rarely accepted (\Cref{sec:baseline}), so most steps stop at the test and skip the costly decoding.
Under the locally balanced proposal, the acceptance of \Cref{eq:zanella_accept_approx} is of order unity, and nearly every step calls the decoder.
Drawing the proposal itself is cheap, since the weights $\lambda_i(e)$ take only two values (\Cref{eq:zanella_weight_sqrt}).
The trade-off is therefore between faster convergence within a sector and many more decoder calls per step, and we quantify it in the time-to-accuracy benchmark of \Cref{sec:results}.

\subsection{Cluster exchange}
\label{sec:cluster}
The locally balanced proposal still moves one bit at a time, so the chain must climb the same high-weight barrier as under \obf{} to reach another sector.
Under both kernels, an accepted move increases the weight with probability of order $(n-w)p/w$ at small $p$ (\Cref{eq:zanella_normalizer}), so a climb of $\Delta w$ levels is suppressed by $p^{\Delta w}$ in either case and the locally balanced chain climbs no faster than \obf{}. The kernels differ only in how many rejected steps separate the accepted moves.

The cluster-exchange move is designed to bypass this barrier by preserving the Hamming weight.
From the current configuration, the move selects a fixed number $m$ of the occupied qubits ($e_i = 1$) and $m$ of the unoccupied qubits ($e_i = 0$), each set uniformly at random, and exchanges their occupancy.
When the configuration has fewer than $m$ occupied or fewer than $m$ unoccupied qubits, no exchange is possible and the chain keeps its current state.
Every admissible candidate differs from the current configuration $e$ in exactly $m$ occupied and $m$ unoccupied positions, all drawn uniformly.
The proposal is therefore
\begin{align}
  q(e'\mid e) = \binom{w(e)}{m}^{-1}\binom{n-w(e)}{m}^{-1}
  \label{eq:cluster_proposal}
\end{align}
for every configuration $e'$ with $w(e') = w(e)$ and $w(e'+e) = 2m$, and $q(e'\mid e) = 0$ otherwise.
When no exchange is possible, the proposal instead puts all its weight on the current state, $q(e\mid e)=1$.

The proposal of \Cref{eq:cluster_proposal} is symmetric, $q(e\mid e') = q(e'\mid e)$, since the equal weights $w(e') = w(e)$ make the two binomial factors coincide.
It also preserves the Hamming weight, so the target probability is unchanged, $\pi_{p}(e') = \pi_{p}(e)$. 
The Metropolis--Hastings ratio of \Cref{eq:mh} then reduces to
\begin{align}
  A(e,e') = \min\!\left(1,\ \frac{\pi_{p}(e')\,q(e\mid e')}{\pi_{p}(e)\,q(e'\mid e)}\right) = 1.
  \label{eq:cluster_accept}
\end{align}
Hence, the move is accepted whenever the proposed configuration $e'$ remains on the fault manifold.
The exchange relocates $m$ errors at fixed weight $w$, so a sequence of exchanges can connect faulty configurations of equal weight in different logical sectors without a weight-increasing move, provided every intermediate configuration is faulty.
It thus complements the locally balanced proposal, which changes the weight.

The cluster-exchange move itself is well known in statistical physics as the nonlocal variant of Kawasaki dynamics at fixed particle number~\cite{kawasaki1966diffusion,newman1999monte}.
What is specific to our setting is that, on the fault manifold, the exchange is accepted or rejected by the decoder rather than by a Boltzmann factor.

The move has two downsides, the cost of the acceptance test and the lack of ergodicity on its own.
First, because the acceptance test of \Cref{alg:mh} is passed trivially by \Cref{eq:cluster_accept}, every step calls the decoder.
The cluster-exchange move thus adds computation time, especially when decoding is expensive, and the computation times reported in \Cref{sec:results} include these extra decoder calls.
Second, because it never changes the weight, cluster exchange cannot reach faulty configurations of other weights, so it is not ergodic by itself and must be combined with a weight-changing move such as the locally balanced one.

\section{Numerical simulation setup}
\label{sec:setup}

To benchmark the proposed kernel, we conduct numerical experiments that measure its performance across codes and decoders.
This section specifies the setup used in each experiment in this paper.
It covers the codes, decoders, and noise model, together with the direct-sampling methods that serve as the competitor and the reference.
It then specifies the two benchmarks, each with its Markov chain parameters, and the conventions for the averages and uncertainties reported in \Cref{sec:results}.
These defaults apply throughout unless otherwise stated.

\subsection{Codes, decoders, and noise model}
\label{sec:setup-codes}
We list the codes, which span four families, and their standard decoders in \Cref{tab:codes}.
\begin{itemize}
\item The topological family comprises the square surface codes $[[d^2+(d-1)^2,1,d]]$ at $d=3$, $5$, $7$ and the square toric codes $[[2d^2,2,d]]$ at $d=5$, $7$, together with the asymmetric toric codes $\mathrm{AT}(d_1,d_2)=[[2d_1d_2,2,d_1]]$, that is, toric codes on a $d_1\times d_2$ torus~\cite{kitaev2003fault,dennis2002topological}, at $(d_1,d_2)=(3,11)$ and $(3,15)$, decoded by minimum-weight perfect matching (\mwpm{})~\cite{higgott2022pymatching}.
\item The concatenated family is the level-$2$ self-concatenated Steane code, $[[49,1,9]]$, decoded by a hard-decision lookup-table decoder.
\item The hypergraph product family is represented by HGP(Hamming) $[[58,16,3]]$ (the hypergraph product code~\cite{tillich2014quantum} from the $[7,4,3]$ Hamming code with itself).
It is decoded by \bposd{}~\cite{panteleev2021degenerate} of the \texttt{ldpc} package, version 2.3.9~\cite{roffe2022ldpc}, with the min-sum belief propagation at scaling factor $1$ under the parallel schedule for $100$ iterations, followed by ordered statistics decoding with the combination-sweep method at order $10$, and with the physical error rate $p$ of the run as the channel prior.
\item The bivariate bicycle (BB) family~\cite{bravyi2024highthreshold} comprises $[[72,12,6]]$, $[[144,12,12]]$, and $[[108,8,10]]$, decoded by \bposd{} with the same settings, following Ref.~\cite{bravyi2024highthreshold}.
The three codes are built on $6\times6$, $12\times6$, and $9\times6$ cyclic lattices with the polynomials $A=x^3+y+y^2$ and $B=y^3+x+x^2$ in the notation of that reference.
\end{itemize}

\begin{table*}[t]
\centering
\small
\begin{tabular}{llccl}
\toprule
family & code & $[[n,k,d]]$ & spread & decoder \\
\midrule
topological  & surface, $d=3,5,7$                    & $[[d^2+(d-1)^2,1,d]]$ & $0$       & \mwpm{} \\
             & toric, $d=5,7$                        & $[[2d^2,2,d]]$        & $0$       & \mwpm{} \\
             & $\mathrm{AT}(3,d_2)$, $d_2=11,15$     & $[[6d_2,2,3]]$        & $d_2-3$   & \mwpm{} \\
\midrule
concatenated & Steane, level $2$                     & $[[49,1,9]]$          & $0$       & lookup table \\
\midrule
HGP          & HGP(Hamming)                          & $[[58,16,3]]$         & $1$       & \bposd{} \\
\midrule
BB &                                       & $[[72,12,6]]$         & $8$       & \bposd{} \\
             &                                       & $[[144,12,12]]$       & $12$      & \bposd{} \\
             &                                       & $[[108,8,10]]$        & $18$      & \bposd{} \\
\bottomrule
\end{tabular}
\caption{Codes and decoders of the main text.
Spread is the basis-weight spread $\max_i w((L_X)_i)-\min_i w((L_X)_i)$ of the logical-$X$ basis used in the experiments.}
\label{tab:codes}
\end{table*}

To characterize the risk of trapping, we consider the basis-weight spread of the logical-$X$ basis.
It is defined as the difference between the maximum and minimum weights of the logical-$X$ basis vectors,
\begin{align}
  \max_i w((L_X)_i)-\min_i w((L_X)_i) .
  \label{eq:basis-weight-spread}
\end{align}
With the standard choice of logical basis, it is zero on the square topological codes and $d_2-d_1$ on the asymmetric toric family $\mathrm{AT}(d_1,d_2)$.
The basis is the one returned by the code construction, namely the logical operators of the \texttt{bposd} implementation of the hypergraph product for the surface, toric, asymmetric toric, and HGP codes, and a row-reduced basis of the kernel of $H_Z$ modulo the row space of $H_X$ for the BB and concatenated Steane codes.
The spread depends on the choice of logical basis rather than on the code alone, so we use it only as a diagnostic.\footnote{The basis enters the experiments only through the initial states of the chains, the maximum- and minimum-weight basis elements of \Cref{sec:bench1}. For an ergodic kernel, the estimate becomes independent of the initial state and unbiased in the long-chain limit, whichever basis element it starts from. Conversely, a bias from the elements of some basis shows that the kernel is trapped on the budget used. We therefore take a basis with a given spread and test whether the estimates from its elements are unbiased.}

Our default noise model is independent and identically distributed bit-flip (Pauli $X$) noise at rate $p$ per qubit.
We repeat the benchmark under depolarizing noise on the [[144,12,12]] BB code in \Cref{app:depol}.

\subsection{Direct-sampling estimates}
\label{sec:setup-decoders}
\label{sec:setup-stratified}
We compare the MCMC methods with direct (rejection) Monte Carlo in the time-to-accuracy benchmark (\Cref{sec:bench2}).
For each code and error rate $p$, each trial draws an error configuration $e$ from the noise model and decodes it once, and we record every faulty draw.
Sampling runs for a fixed budget of computation time, $96$ single-core runs of $300$ seconds for each code and error rate, and we pool the faulty draws of all runs.
Within this budget, the sampler makes of order $10^{7}$ to $10^{10}$ trials, depending on the code and the error rate.
The number of faulty draws collected falls steeply with the error rate.
At $p=10^{-1}$, it is of order $10^{7}$ to $10^{8}$.
At $p=10^{-3}$, it ranges from a single draw to $10^{6}$.
At $p=10^{-5}$, it is at most $10^{2}$, and eight of the twelve codes yield none.

We compute the direct Monte Carlo values and their standard errors as follows.
Let $N_{\mathrm{trial}}$ be the number of trials for a given code and error rate $p$, $N_{\mathrm{faulty}}$ the number of faulty draws $e^{(i)}$ among them, and $\mathcal{O}(e)$ an observable on the fault manifold, which throughout this paper is the weight $w(e)$.
The values are estimated as
\begin{align}
\hat P_L &\coloneqq\frac{N_{\mathrm{faulty}}}{N_{\mathrm{trial}}},
\label{eq:dmc-pl} \\
\hat{\mathbb{E}}_{\mathcal{F}}[\mathcal{O}] &\coloneqq\frac{1}{N_{\mathrm{faulty}}}\sum_{i=1}^{N_{\mathrm{faulty}}} \mathcal{O}(e^{(i)}),
\label{eq:dmc-expect}
\end{align}
and their standard errors are estimated as
\begin{align}
  \label{eq:dmc-se}
\sigma(\hat P_L)&\coloneqq\sqrt{\frac{\hat P_L(1-\hat P_L)}{N_{\mathrm{trial}}-1}}\simeq\frac{\hat P_L}{\sqrt{N_{\mathrm{faulty}}}},
\\ 
\sigma\bigl(\hat{\mathbb{E}}_{\mathcal{F}}[\mathcal{O}]\bigr)&\coloneqq\sqrt{\frac{\widehat{\mathrm{Var}}_{\mathrm{ideal}}}{N_{\mathrm{faulty}}}},
\end{align}
where the approximation is valid for $P_L\ll1$.
Here, $\widehat{\mathrm{Var}}_{\mathrm{ideal}}$ is the unbiased sample variance of the observed values $\mathcal{O}(e^{(i)})$, which estimates the variance of the observable over the fault manifold (the ideal variance of \Cref{sec:bench1}).
Here and below, a hat marks a quantity estimated from samples.
The sample variance is given by
\begin{align}
  \label{eq:sample-variance}
\widehat{\mathrm{Var}}_{\mathrm{ideal}}&\coloneqq\frac{1}{N_{\mathrm{faulty}}-1}\sum_{i=1}^{N_{\mathrm{faulty}}}\bigl(\mathcal{O}(e^{(i)})-\hat{\mathbb{E}}_{\mathcal{F}}[\mathcal{O}]\bigr)^2.
\end{align}

One implementation detail affects the computation times quoted later.
The decoders are deterministic functions of the syndrome.
This allows every process to save computation time by keeping a cache of the corrections keyed by syndrome and calling the decoder only on a syndrome that is not in the cache.
The cache is cleared once it holds $2\times10^{5}$ entries, to limit the memory usage.
The same procedure is used for the MCMC kernels of \Cref{sec:bench1}.

To judge whether an MCMC estimate is correct, we also need reference values obtained without a Markov chain.
We use two direct-sampling methods for the reference, namely direct Monte Carlo and the weight-stratified estimate described below.
Direct Monte Carlo is available only at high error rates, so we use the weight-stratified estimate at low error rates.

The weight-stratified estimate, known as importance sampling over the error weight~\cite{beverland2025failfast} or subset sampling~\cite{heussen2024subset}, draws uniformly at a fixed weight rather than from the noise distribution $\pi_p(e)$.
Under i.i.d.\ bit-flip noise at rate $p$ on $n$ qubits, the weight $w(e)$ is binomially distributed.
Splitting the fault manifold $\mathcal{F}$ into \emph{weight classes}, the sets of faulty configurations of a given weight, therefore gives
\begin{align}
  P_L&=\sum_{w} B(w;n,p)\,f_w,
  \label{eq:stratified}
  \\
  \mathbb{E}_{\mathcal{F}}[w]&=\frac{\sum_{w} w\,B(w;n,p)\,f_w}{\sum_{w} B(w;n,p)\,f_w},
  \label{eq:stratified-ew}
\end{align}
where $B(w;n,p)\coloneqq\binom{n}{w}p^{w}(1-p)^{n-w}$ is the binomial weight distribution and $f_w\coloneqq\Pr(e\in\mathcal{F}\mid w(e)=w)$ is the failure fraction of the weight class of weight $w$.

The failure fraction depends on the code and its decoder alone, and not on the error rate, except for \bposd{}, which takes the error rate as an input (\Cref{sec:decoding}).
Its failure fractions on the BB codes change substantially with the error rate, so we recompute them at each error rate $p$.
On the $[[144,12,12]]$ BB code, for example, the decoder fails on some configurations of weight $3$ at every error rate, which is why the mean fault weight of this distance-$12$ code is close to $3$ in \Cref{tab:meanw}.

We obtain the failure fraction $f_w$ of a weight class by decoding every error in the class for small weights $w$, where the class can be enumerated, and by decoding errors drawn uniformly at random from the class for larger weights.
The standard error of a sampled failure fraction is the Agresti--Coull estimate, and the standard errors of the logical error rate $P_L$ and of the mean weight follow from those of the independently sampled weight classes by the delta method.
These relative errors calculated by this method are below $10^{-3}$ and are negligible.

The cost is set by the weight classes rather than by the logical error rate, so the estimate remains affordable down to $p=10^{-5}$.
We truncate the sums in \Cref{eq:stratified,eq:stratified-ew} at a maximum weight $w_{\max}$ ($16$ for \mwpm{} and the lookup table, $12$ for \bposd{}), which neglects a probability of order $10^{-9}$ for $p\le10^{-2}$. We use the estimate from this sampling as the MCMC-independent reference in that range.
At $p=10^{-1}$, the direct Monte Carlo value is available, and thus we use it as the reference.

The failure fraction $f_w$ can be measured only for a weight class that is small enough to enumerate or in which failures are frequent enough to be collected by uniform sampling.
On our benchmark codes, every weight class up to the cutoff weight $w_{\max}$ meets one of the two conditions.
At small error rates, however, the sums of \Cref{eq:stratified,eq:stratified-ew} are dominated by the classes near the minimum fault weight, which grows with the code distance, and on a code of large distance these classes are too large to enumerate while their failures are too rare to collect by uniform sampling.
The weight-stratified estimate is therefore a reference for codes of moderate distance, such as our benchmark codes, and not a substitute for MCMC on larger codes.

\subsection{First benchmark, accuracy of the mean fault weight}
\label{sec:bench1}

We conduct two numerical benchmarks of the kernels, both on the mean fault weight $\mathbb{E}_{\mathcal{F}}[w]$.
The first tests the accuracy of each kernel's estimate against an MCMC-independent reference at a fixed budget, and the second measures the time to reach a target accuracy.
This subsection specifies the setup of the first, and \Cref{sec:bench2} that of the second.

The first benchmark checks the mean fault weight $\mathbb{E}_{\mathcal{F}}[w]$ of \Cref{eq:meanw}, the expectation of the Hamming weight $w(e)$ over the fault manifold $\mathcal{F}$.
We compute the MCMC estimate $\hat{\mathbb{E}}_{\mathcal{F}}[w]$ and its standard error from the chains as in \Cref{eq:chain-estimate,eq:chain-se} below with the observable $\mathcal{O}=w$, and compare them with the MCMC-independent reference of \Cref{sec:setup-stratified}, the weight-stratified value of \Cref{eq:stratified-ew}.

This benchmark aims to measure directly how accurately each kernel samples the fault manifold within a fixed budget.
We check whether each kernel returns the correct mean fault weight by comparing the estimate with the independent reference.
In addition, we apply the convergence rule defined below, which is built only from the statistics of the chains themselves and does not use the reference.
The rule therefore remains available in practice, where no reference exists.

As noted in \Cref{sec:baseline}, a kernel that is not ergodic, or ergodic but slow to cross between logical sectors within a fixed budget, returns an estimate that depends on the initial state.
Both checks above can detect this failure.
The comparison with the reference shows the bias directly, and the convergence rule flags an estimate whose chains have not settled or whose value depends on the initial state.

We generate every Markov chain with the Metropolis--Hastings algorithm of \Cref{alg:mh}.
Each case, one kernel on one code at one error rate, uses the following setup unless stated otherwise.
\begin{itemize}
\item Every case runs $50$ independent chains of $N=2\times10^{6}$ steps each.
\item Each chain starts from one of two deliberately separated states, the maximum-weight (long) or the minimum-weight (short) logical-$X$ basis element.

\item We compare the default kernel \zh{}, the locally balanced proposal of \Cref{sec:zanella} combined with the cluster exchange of \Cref{sec:cluster}, the \obf{} baseline.
It runs at cluster-exchange probability $p_{\mathrm{cluster}}=0.3$ and exchange size $m=2$.
\item To remove the initial transient, we discard the first $5\iat$ steps of each chain (burn-in~\cite{newman1999monte,geyer2011handbook}), with $\iat$ the estimate of the integrated autocorrelation time (\Cref{eq:iat,eq:iat-hat} below), re-estimated after each burn-in cut until it is stable.
Where that estimate does not converge, as on slow or trapped chains, the burn-in length is set to half the total chain length.
The steps that remain after the burn-in, $N'$ per chain, are the samples used below.
Since the burn-in is set chain by chain, $N'$ differs between chains, and the convergence rule below uses its median over the chains.
We evaluate the chain means, the within-chain variances, and $\iat$ on every tenth step of these chains, and quote $\iat$ and $N'$ in units of the original steps.
\end{itemize}

We first fix how we estimate the mean of an observable and its standard error, and then state the convergence rule, which is built from these quantities.
Since successive samples of a Markov chain are correlated, the standard error of an MCMC estimate is not given by the formula of \Cref{eq:dmc-se} for independent samples.
In this work, we therefore take the statistical errors from the spread between independent chains.
We run $K=50$ independent chains and denote by $e^{(k)}_1,\dots,e^{(k)}_{N'}$ the configurations that chain $k$ visits after the burn-in.
For an observable $\mathcal{O}$, the chain mean of chain $k$ is
\begin{align}
\bar{\mathcal{O}}_k\coloneqq\frac{1}{N'}\sum_{s=1}^{N'}\mathcal{O}\bigl(e^{(k)}_s\bigr).
\label{eq:chain-mean}
\end{align}
The average of the chain means $\bar{\mathcal{O}}_1,\dots,\bar{\mathcal{O}}_K$ and its standard error are given by
\begin{align}
\hat{\mathbb{E}}_{\mathcal{F}}[\mathcal{O}] &\coloneqq\frac{1}{K}\sum_{k=1}^{K}\bar{\mathcal{O}}_k,
\label{eq:chain-estimate} \\
\sigma\bigl(\hat{\mathbb{E}}_{\mathcal{F}}[\mathcal{O}]\bigr) &\coloneqq\sqrt{\frac{\widehat{\mathrm{Var}}_{\mathrm{cross}}}{K}},
\label{eq:chain-se}
\end{align}
where the cross-chain variance $\widehat{\mathrm{Var}}_{\mathrm{cross}}$ is defined as
\begin{align}
\widehat{\mathrm{Var}}_{\mathrm{cross}} \coloneqq \frac{1}{K-1} \sum_{k=1}^{K} \left(\bar{\mathcal{O}}_k - \hat{\mathbb{E}}_{\mathcal{F}}[\mathcal{O}] \right)^2,
\label{eq:chain-variance}
\end{align}
that is, the sample variance of the chain means $\{\bar{\mathcal{O}}_k\}_{k=1}^{K}$ as in \Cref{eq:sample-variance}.

We now relate the cross-chain variance to two other variances, the single-chain variance of one chain mean and the ideal variance of the observable over the fault manifold $\mathcal{F}$.
Let $e_1,\dots,e_{N'}$ be the samples of one chain after the burn-in, and assume that the chain is stationary after the burn-in, so that each state $e_s$ is distributed according to the conditional fault distribution $\pi_{p\mid\mathcal{F}}$.
Let $\bar{\mathcal{O}}$ be its chain mean and $\mathrm{Var}_{\mathrm{single}}$ the single-chain variance, the variance of that mean over realizations of the chain, defined as
\begin{align}
  \bar{\mathcal{O}}&\coloneqq \frac{1}{N'}\sum_{s=1}^{N'}\mathcal{O}(e_s), \\
  \mathrm{Var}_{\mathrm{single}}
  & \coloneqq \mathbb{E}\Bigl[\bigl(\bar{\mathcal{O}}-\mathbb{E}_{\mathcal{F}}[\mathcal{O}]\bigr)^2\Bigr] \\
  &= \frac{\mathrm{Var}_{\mathrm{ideal}}}{N'}\left(1+2\sum_{t=1}^{N'-1}\left(1-\frac{t}{N'}\right)\rho(t)\right),
\end{align}
where $\mathrm{Var}_{\mathrm{ideal}}\coloneqq\mathbb{E}_{\mathcal{F}}[\mathcal{O}^2]-\mathbb{E}_{\mathcal{F}}[\mathcal{O}]^2$ is the ideal variance, the variance of the observable over the fault manifold, and
\begin{align}
  \rho(t)\coloneqq\frac{\mathbb{E}[\mathcal{O}(e_s)\,\mathcal{O}(e_{s+t})]-\mathbb{E}_{\mathcal{F}}[\mathcal{O}]^2}{\mathrm{Var}_{\mathrm{ideal}}}
\end{align}
is the normalized autocorrelation at lag $t$, which does not depend on the starting step $s$ for a stationary chain.
The integrated autocorrelation time is defined as
\begin{align}
\tau\coloneqq1+2\sum_{t=1}^{\infty}\rho(t),
\label{eq:iat}
\end{align}
which is the number of successive steps that carry the information of one independent sample.
When the number of steps $N'$ is sufficiently large, the variance of the chain mean can be approximated as
\begin{align}
\mathrm{Var}_{\mathrm{single}}\simeq\frac{\tau}{N'}\,\mathrm{Var}_{\mathrm{ideal}}.
\label{eq:chain-var}
\end{align}
The single-chain variance is thus larger by the factor $\tau$ than the ideal variance.

In practice, the autocorrelation time $\tau$ has to be estimated from the chain itself.
We estimate it by replacing the exact autocorrelation $\rho(t)$ in \Cref{eq:iat} by the empirical autocorrelation of the samples, following the initial-positive-sequence idea of Geyer~\cite{geyer1992practical}, and denote the estimate by $\iat$, defined as
\begin{align}
  \iat&\coloneqq1+2\sum_{t=1}^{t_{\mathrm{c}}-1}\hat\rho(t),
  \label{eq:iat-hat}
  \\
  \hat\rho(t)&\coloneqq\frac{\sum_{s=1}^{N'-t}\bigl(\mathcal{O}(e_s)-\bar{\mathcal{O}}\bigr)\bigl(\mathcal{O}(e_{s+t})-\bar{\mathcal{O}}\bigr)}{\sum_{s=1}^{N'}\bigl(\mathcal{O}(e_s)-\bar{\mathcal{O}}\bigr)^2},
  \label{eq:rho-hat}
\end{align}
where $t_{\mathrm{c}}$ is the first lag at which $\hat\rho(t)$ is negative.
This estimate is meaningful only if the chain is stationary and the autocorrelation has decayed within the chain, although neither can be verified from the samples alone.
When the chain has not yet explored the whole fault manifold, the estimate $\iat$ measures only the decorrelation within the region visited so far and underestimates the true uncertainty.
We therefore compute the statistical errors not from this estimate but from the cross-chain variance $\widehat{\mathrm{Var}}_{\mathrm{cross}}$.

The $K$ chains are independent, and if the burn-in has brought them to stationarity, the chain means $\bar{\mathcal{O}}_1,\dots,\bar{\mathcal{O}}_K$ are independent random variables with mean $\mathbb{E}_{\mathcal{F}}[\mathcal{O}]$ and variance $\mathrm{Var}_{\mathrm{single}}$.
The cross-chain variance $\widehat{\mathrm{Var}}_{\mathrm{cross}}$ therefore estimates the single-chain variance $\mathrm{Var}_{\mathrm{single}}$, and its value divided by the number of chains, $\widehat{\mathrm{Var}}_{\mathrm{cross}}/K$, estimates the variance of the estimate $\hat{\mathbb{E}}_{\mathcal{F}}[\mathcal{O}]$, which gives the standard error in \Cref{eq:chain-se}.
Since $\widehat{\mathrm{Var}}_{\mathrm{cross}}$ is computed from independent chains, \Cref{eq:chain-se} is valid under residual autocorrelation within each chain and needs no estimate of the autocorrelation time.

The convergence rule below also uses the within-chain variance
\begin{align}
  \widehat{\mathrm{Var}}_{\mathrm{within}}\coloneqq\frac{1}{K}\sum_{k=1}^{K}\frac{1}{N'-1}\sum_{s=1}^{N'}\bigl(\mathcal{O}(e^{(k)}_s)-\bar{\mathcal{O}}_k\bigr)^2,
  \label{eq:within-variance}
\end{align}
which is the $W$ of the Gelman--Rubin statistic~\cite{gelman1992inference}.
If the chain is stationary, each of its states is a sample from the conditional fault distribution $\pi_{p\mid\mathcal{F}}$, so the spread of the states within one chain estimates the ideal variance, and the within-chain variance $\widehat{\mathrm{Var}}_{\mathrm{within}}$ converges to the ideal variance $\mathrm{Var}_{\mathrm{ideal}}$ as the number $N'$ of samples per chain goes to infinity, for any number $K$ of chains.

The convergence rule is a separate test applied after the run.
It decides which entries of \Cref{sec:results} we report as converged.
All the following four parts must pass.
\begin{enumerate}[label=(\roman*)]
\item The cross-chain standard deviation of the chain means shrinks from the first $N/4$ steps to all $N$ steps at least as fast as the inverse square root of the length within a factor $1.5$, that is, $2\sqrt{\widehat{\mathrm{Var}}_{\mathrm{cross}}(N)/\widehat{\mathrm{Var}}_{\mathrm{cross}}(N/4)}\le1.5$, where $\widehat{\mathrm{Var}}_{\mathrm{cross}}(N/4)$ and $\widehat{\mathrm{Var}}_{\mathrm{cross}}(N)$ are the cross-chain variances of the chain means over the first $N/4$ and over all $N$ steps.
\item The average of the chain means over the last half of the steps and that over the preceding quarter differ by no more than three standard errors of their difference (the first quarter is dropped to exclude the transient from a distant start).
The standard error of the difference is $\sqrt{6}$ times the standard error $\sigma$ of \Cref{eq:chain-se} over all $N$ steps.
This factor assumes that the variance of a segment mean scales inversely with its length, so that the means over the last half and over the preceding quarter have variances $2\sigma^{2}$ and $4\sigma^{2}$, and that the two segments are independent, so that the variance of the difference is their sum $6\sigma^{2}$.
\item Where a case is run from two initial states, its two estimates differ by no more than three standard errors of their difference.
\item The cross-chain standard deviation $\sqrt{\widehat{\mathrm{Var}}_{\mathrm{cross}}}$ does not exceed three times $\sqrt{\widehat{\mathrm{Var}}_{\mathrm{within}}\,\iat/N'}$, its value predicted by \Cref{eq:chain-var} with the within-chain variance of \Cref{eq:within-variance} in place of the ideal variance.
Here $\iat$ and $N'$ are the medians over the chains.
\end{enumerate}
Each part detects a different failure.
Part (i) checks the approximate inverse-square-root scaling with the length that is expected for stationary, independently seeded chains once the chains are long compared with the autocorrelation time. Chains that have not yet reached stationarity may violate it even while their mean looks stable.
Part (ii) catches a mean that has not settled by the end of the run.
Part (iii) compares the estimates from the two initial states.
It is the only part that detects chains trapped together in one logical sector.
Part (iv) checks the observed cross-chain variance against the value the autocorrelation time predicts.
Note that passing all four parts does not prove that an estimate is unbiased, since the rule detects only failures that show up in the statistics of the chains.

\subsection{Second benchmark, time to accuracy}
\label{sec:bench2}

The second benchmark quantifies the \emph{time to accuracy} of each kernel, and of direct Monte Carlo, in total single-core-equivalent computation time.
All computation times in this paper are recorded on single cores of cluster nodes with two $48$-core Intel Xeon Platinum 9242 processors at $2.3$ GHz, running Python 3.11 with the package versions listed in the repository of the Data availability statement.
The time to accuracy is the computation time until the estimate of the mean fault weight comes within $2\%$ of the reference and stays there.

To find it, we follow the estimate as the chains grow.
We record each chain at the checkpoints $L=10^{3}\cdot2^{j}$ ($j=0,1,2,\dots$) steps and apply \Cref{eq:chain-estimate,eq:chain-se} with the observable $\mathcal{O}=w$ to the first $L$ steps of each chain, which gives the estimate $\hat{\mathbb{E}}_L$ and its standard error $\sigma_L$ at checkpoint $L$.
We then measure the error of $\hat{\mathbb{E}}_L$ against the reference value $\mathbb{E}_{\mathrm{ref}}$ from the direct-sampling methods of \Cref{sec:setup-stratified}.
The relative error at checkpoint $L$ is defined as
\begin{align}
\epsilon(L)\coloneqq\frac{\max\bigl(\bigl|\hat{\mathbb{E}}_L-\mathbb{E}_{\mathrm{ref}}\bigr|,\ \sigma_L\bigr)}{\mathbb{E}_{\mathrm{ref}}}.
\label{eq:rel-rmse}
\end{align}
We require the standard error $\sigma_L$ to be below the target, as for direct Monte Carlo below.
We also require the deviation from the reference to be below the target, since a Markov chain can be biased.
The relative standard error of the reference is at most $8\times10^{-4}$, which is negligible against the $2\%$ target in this benchmark.

Unlike in the first benchmark, we do not discard a burn-in here.
The burn-in there is set by the autocorrelation time estimated from the chain itself, and the first $L$ steps do not yet give a reliable estimate of the autocorrelation time at the early checkpoints.
Rather than adapting the burn-in to each kernel and code, we compare all cases under a single rule by using all steps so far at every checkpoint.
A burn-in, if used, would mainly remove the transient bias, so omitting it is expected to make the speedups of \zh{} over direct Monte Carlo conservative.
Without a burn-in, the initial transient enters $\hat{\mathbb{E}}_L$ as a bias that decays as $1/L$, and hence enters the deviation in \Cref{eq:rel-rmse}.

We set the target to $\epsilon(L)<2\%$.
The time to accuracy is the computation time at the earliest checkpoint $L^{*}$ at which the error $\epsilon$ is below the target and stays below it at every later checkpoint,
\begin{align}
T_{2\%}&\coloneqq t\bigl(L^{*}\bigr),
\label{eq:tta}
\\
L^{*}&\coloneqq\min\{L:\ \epsilon(L')<0.02\ \text{for all}\ L'\geq L\},
\label{eq:tta-lstar}
\end{align}
which excludes an optimistic first crossing by chance.
A crossing only at the largest recorded checkpoint counts as not reached, since no later checkpoint confirms it.
Here, the computation time $t(L)$ is the total time consumed up to checkpoint $L$, that is, the sum over the $K$ chains of the wall-clock time each of them spent reaching the checkpoint.
The error $\epsilon(L)$ is that of the average over all $K$ chains, so we count the computation time of all $K$ chains, not of a single one.

For comparison, we also measure the time to accuracy of direct Monte Carlo.
Its estimate carries no bias, so its error $\epsilon$ is the relative standard error $\sigma(\hat{\mathbb{E}}_{\mathcal{F}}[w])/\hat{\mathbb{E}}_{\mathcal{F}}[w]$ of \Cref{eq:dmc-se}. Its time to accuracy $T_{2\%}$ is the number of core-seconds needed to collect at least one faulty draw and to bring the error below $2\%$.
The variance in \Cref{eq:dmc-se} is taken from the reference computation of \Cref{sec:setup-stratified} rather than from the run itself, so the required number of faulty draws is fixed by the code and the error rate.
We obtain it from the rate of faulty draws in the fixed-budget run of \Cref{sec:setup-decoders}.
Where that run collected no faulty draw, we bound its rate from above by one draw per budget, and where it collected some but fewer than needed, we charge the budget itself, so the quoted time is a lower bound in both cases.

The time-to-accuracy benchmark measures how quickly each kernel reaches the $2\%$ target on the mean fault weight.
It compares the computational efficiency of the kernels on an equal footing, since the cost counts both the initial transient and the chain length required.
As discussed in \Cref{sec:kernel}, \zh{} is designed to explore the fault manifold efficiently, but its cluster-exchange move costs one decoder call per proposal, and this overhead may offset the gain in mixing.
Even for a kernel that mixes well in principle, the relevant quantity is the computation time it takes to reach the target.

We also note that the convergence rules of the two benchmarks differ in kind.
In the first benchmark, the criterion for convergence relies only on the output of the chains, since the convergence rule of \Cref{sec:bench1} does not use the reference.
The rule checks whether the estimate depends on the initial state and whether the behavior of its standard error is consistent with ergodicity.

In the second benchmark, in contrast, the criterion for accuracy relies on the reference as well, since the error $\epsilon$ of \Cref{eq:rel-rmse} is defined with respect to the reference value.
The benchmark checks whether the estimate reaches the $2\%$ target within the budget and how much computation time it needs to do so.
If the bias of a kernel exceeds $2\%$ of the reference $\mathbb{E}_{\mathrm{ref}}$, the error $\epsilon(L)$ stays above the target however small the standard error $\sigma_L$ becomes, and the kernel never reaches the $2\%$ target.
The time to accuracy $T_{2\%}$ thus measures the cost of an estimate that is within $2\%$ of the reference.
We therefore do not apply the convergence rule of \Cref{sec:bench1} in this benchmark.

As in \Cref{sec:bench1}, we generate every Markov chain with the Metropolis--Hastings algorithm of \Cref{alg:mh}.
Each case uses the following setup.
\begin{itemize}
\item Every case runs $K=50$ independent chains, as in \Cref{sec:bench1}.
The chains run with checkpoints up to $L=10^{3}\cdot2^{15}\approx3.3\times10^{7}$.
For \obf{} on the $[[144,12,12]]$ BB code at $p=10^{-3}$ and $10^{-5}$, the checkpoints extend to $L=10^{3}\cdot2^{18}\approx2.6\times10^{8}$.
\item All chains of both kernels start from the same state, the maximum-weight (long) initialization. This is in contrast to \Cref{sec:bench1}, which also uses the minimum-weight (short) one.
The long start is the worse one for \obf{} on the asymmetric toric codes (\Cref{tab:meanw}), so their \obf{} entries in \Cref{tab:obfspeedup} include the cost of a wrong start. We keep it because the convergence rule of \Cref{sec:discussion} applied to a new code needs chains from a far start, and those chains dominate the computation time.
\item We compare the default kernel \zh{}, run with the same parameters as in \Cref{sec:bench1}, with the \obf{} baseline.
\item We do not use the burn-in, unlike in \Cref{sec:bench1}.
\end{itemize}

We repeat the whole benchmark ten times with disjoint seed streams and report, for each method, the mean of the time to accuracy $T_{2\%}$ over the ten runs.
We report the speedup of \zh{} over each competitor, direct Monte Carlo or \obf{}.
The speedup is the mean time to accuracy of the competitor divided by that of \zh{}, so a speedup above unity means that \zh{} is faster.

We estimate the standard error of the speedup by the bootstrap with $2\times10^{4}$ resamples over the ten runs.
If a method did not reach the target within its budget in at least one of the ten runs, its mean time and the speedup built on it are lower bounds, marked with the inequality sign $\geq$ in the tables.

\section{Results}
\label{sec:results}

This section presents the results of the two benchmarks of \Cref{sec:bench1,sec:bench2}.
\Cref{tab:meanw} compares the mean fault weight returned by each kernel with the reference on twelve codes at the fixed physical error rate $p=10^{-4}$, and \Cref{tab:lowp} follows the two broad-spread BB codes across error rates.
\Cref{tab:dmcspeedup} gives the time-to-accuracy speedup of \zh{} over direct Monte Carlo, and \Cref{tab:obfspeedup} that over the \obf{} baseline.
We discuss them in that order.

\subsection{Accuracy benchmarks}

\begin{table*}[t]
\centering
\small
\setlength{\tabcolsep}{2.5pt}
\begin{tabular}{lcccccc}
\toprule
 & & & \multicolumn{2}{c}{\obf{} estimated bias} & \multicolumn{2}{c}{\zh{} estimated bias} \\
\cmidrule(lr){4-5}\cmidrule(lr){6-7}
code & spread & reference & short & long & short & long \\
\midrule
surface $d=3$                    & $0$  & $2.0006$     & \multicolumn{2}{c}{$-0.000063(10)$} & \multicolumn{2}{c}{$+0.000005(8)$} \\
concat.\ Steane, level $2$       & $0$  & $4.0036$     & \multicolumn{2}{c}{$0.00000(6)$} & \multicolumn{2}{c}{$+0.000012(21)$} \\
\midrule
toric $d=5$                      & $0$  & $3.0082$     & \multicolumn{2}{c}{$-0.00367(6)$} & \multicolumn{2}{c}{$+0.00035(10)$} \\
toric $d=7$                      & $0$  & $4.0174(4)$  & \multicolumn{2}{c}{$\mathbf{-0.0058(18)}^\dagger$} & \multicolumn{2}{c}{$0.0000(6)$} \\
surface $d=5$                    & $0$  & $3.0040$     & \multicolumn{2}{c}{$+0.0005(17)^\dagger$} & \multicolumn{2}{c}{$-0.00002(4)$} \\
surface $d=7$                    & $0$  & $4.0096$     & \multicolumn{2}{c}{$-0.00303(33)$} & \multicolumn{2}{c}{$+0.00005(28)$} \\
\midrule
HGP(Hamming) $[[58,16,3]]$       & $1$  & $2.0061$     & $+0.0003(9)^\dagger$ & $-0.00093(7)^\dagger$ & $+0.000011(29)$ & $+0.000003(28)$ \\
\midrule
$\mathrm{AT}(3,11)$              & $8$  & $2.0073$     & $-0.00101(7)^\dagger$ & $\mathbf{+3.62(16)}^\dagger$ & $-0.00003(4)$ & $0.00000(5)$ \\
$\mathrm{AT}(3,15)$              & $12$ & $2.0097$     & $-0.00108(10)^\dagger$ & $\mathbf{+5.39(24)}^\dagger$ & $-0.00002(5)$ & $-0.00003(6)$ \\
\midrule
BB $[[144,12,12]]$               & $12$ & $3.0548(13)$ & $\mathbf{+0.69(6)}^\dagger$ & $\mathbf{+0.74(6)}^\dagger$ & $+0.0026(21)$ & $+0.0015(22)$ \\
BB $[[108,8,10]]$                & $18$ & $3.0350(8)$  & $\mathbf{+0.30(6)}^\dagger$ & $\mathbf{+0.20(5)}^\dagger$ & $-0.0010(9)$ & $-0.0011(9)$ \\
BB $[[72,12,6]]$                 & $8$  & $3.0125$     & $0.000(4)^\dagger$ & $0.000(6)^\dagger$ & $-0.00005(10)$ & $-0.00005(10)$ \\
\bottomrule
\end{tabular}
\caption{Comparison of mean fault weights across the codes at $p=10^{-4}$ under the setup of \Cref{sec:bench1}.
Parentheses give standard errors. For a bias, the standard error is the chain and reference standard errors added in quadrature.
The reference column is the weight-stratified reference of \Cref{sec:setup-stratified}.
Entries give the estimated signed bias $\hat{\mathbb{E}}_{\mathcal{F}}[w]-\mathbb{E}_{\mathrm{ref}}$ from either the minimum-weight (short) or the maximum-weight (long) initialization.
Zero-spread codes have a single initialization and hence their entries span both columns.
An entry that fails the convergence rule of \Cref{sec:bench1} is marked with a dagger.
Bold marks a bias larger than $3\sigma$ and at least $0.005$ in absolute value. The cutoff of $0.005$ is at most $0.25\%$ of the mean weights in the table, an order of magnitude below the $2\%$ target.
The reference is shown to four decimals, and its standard error is omitted where it is smaller than the last digit.}
\label{tab:meanw}
\end{table*}

We begin with the mean weight across the codes at the fixed physical error rate $p=10^{-4}$, given in \Cref{tab:meanw}.
We compare the MCMC estimates of the baseline \obf{} kernel and of our proposed \zh{} kernel against the weight-stratified reference of \Cref{sec:setup-stratified}.

On the six zero-spread codes, comprising the $k=1$ surface $d=3$ and concatenated Steane codes and the surface and toric codes at $d=5,7$, the \obf{} and \zh{} estimates stay within $0.15\%$ and $0.02\%$ of the reference, respectively, far below the $2\%$ target.

On all other codes, the spread is nonzero, so a chain can start from a minimum-weight (short) or a maximum-weight (long) logical basis element.
It is not known in advance whether \obf{} gives the same estimate from either start, since no proof of ergodicity exists for these codes.

On the HGP(Hamming) code, the \obf{} and \zh{} estimates stay within $0.05\%$ and $0.001\%$ of the reference, respectively, from either start.
This indicates that a nonzero spread alone does not imply trapping.

On the asymmetric toric codes, the \obf{} estimate depends on the initial state.
\obf{} stays within $0.06\%$ of the reference from the short start, whereas it returns a biased estimate from the long start.
Similar behavior has been observed by Beverland et al.~\cite{beverland2025failfast}.
Of the $50$ long-start chains, $45$ on $\mathrm{AT}(3,11)$ and $44$ on $\mathrm{AT}(3,15)$ stay in the high-weight logical sector and return the mean weight of that sector, not of the whole fault manifold $\mathcal{F}$, while the remaining chains escape to the low-weight sector within the budget.
\zh{}, by contrast, agrees with the reference from either start, which shows that it mitigates the trapping of \obf{}.

On the broad-spread BB codes, the trapping of \obf{} is severe.
In the $[[144,12,12]]$ and $[[108,8,10]]$ BB codes, \obf{} is biased from both starts.
The $[[72,12,6]]$ BB code, whose spread of $8$ is the smallest among the three BB codes, shows no \obf{} bias beyond $3\sigma$ from either start, but both of its \obf{} entries fail the convergence rule, with standard errors far larger than those of \zh{}.
\zh{} agrees with the reference from both starts on all three codes.

Biases larger than $0.1$ occur only on the codes with broad spread, and within the asymmetric-toric family, at fixed $k=2$ and fixed decoder, the bias grows with the spread.
The bias also appears on the broad-spread BB codes, and the absolute value of the bias on the $k=2$ asymmetric codes is larger than that on the $[[144,12,12]]$ BB code with $k=12$.
These comparisons associate the bias with the weight spread rather than with the number $k$ of logical qubits.
This is consistent with the conjecture of \Cref{sec:baseline} that codes with broad spread have high-weight barriers between low-weight faulty configurations in different logical sectors, which a one-bit-flip chain crosses at a rate suppressed by powers of the error rate $p$.
Across families, the size of the bias is not monotonic in the spread, so a broad spread tells us that \obf{} may be biased, but not how large the bias is.

\begin{table*}[t]
\centering
\footnotesize
\setlength{\tabcolsep}{3pt}
\begin{tabular}{llccccc}
\toprule
 & & & \multicolumn{2}{c}{\obf{} estimated bias} & \multicolumn{2}{c}{\zh{} estimated bias} \\
\cmidrule(lr){4-5}\cmidrule(lr){6-7}
code & $p$ & reference & short & long & short & long \\
\midrule

BB $[[144,12,12]]$     & $10^{-2}$                  & $4.592(9)$ & $\mathbf{+1.42(12)}^\dagger$ & $\mathbf{+1.64(11)}^\dagger$ & $-0.02(7)$ & $-0.07(8)$ \\
                       & $10^{-3}$                  & $3.0763(17)$ & $\mathbf{+2.47(8)}^\dagger$ & $\mathbf{+2.52(8)}^\dagger$ & $+0.027(29)^\dagger$ & $+0.009(23)^\dagger$ \\
                       & $10^{-4}$                  & $3.0548(13)$ & $\mathbf{+0.69(6)}^\dagger$ & $\mathbf{+0.74(6)}^\dagger$ & $+0.0026(21)$ & $+0.0015(22)$ \\
\midrule
BB $[[108,8,10]]$      & $10^{-2}$                  & $4.961(9)$ & $\mathbf{+0.46(9)}^\dagger$ & $\mathbf{+0.39(8)}^\dagger$ & $+0.029(32)$ & $\mathbf{+0.095(26)}$ \\
                       & $10^{-3}$                  & $3.1604(23)$ & $\mathbf{+1.37(8)}^\dagger$ & $\mathbf{+1.65(9)}^\dagger$ & $-0.025(22)$ & $-0.014(26)$ \\
                       & $10^{-4}$                  & $3.0350(8)$ & $\mathbf{+0.30(6)}^\dagger$ & $\mathbf{+0.20(5)}^\dagger$ & $-0.0010(9)$ & $-0.0011(9)$ \\
\bottomrule
\end{tabular}
\caption{The $[[144,12,12]]$ and $[[108,8,10]]$ BB codes in the rare-event regime.
The reference column is the weight-stratified reference of \Cref{sec:setup-stratified} for $\mathbb{E}_{\mathcal{F}}[w]$.
The kernel columns give the estimated signed bias of the estimate of \Cref{eq:chain-estimate} from each initialization.
Daggers and bold as in \Cref{tab:meanw}.}
\label{tab:lowp}
\end{table*}

We next follow the two broad-spread BB codes, on which \obf{} is biased from both starts (\Cref{tab:meanw}), across error rates, from $p=10^{-2}$ down to $10^{-4}$ in \Cref{tab:lowp}.
The large \obf{} bias persists at every tested error rate and from both starts.
The size of its bias at this chain length depends on the error rate, and it does so nonmonotonically, with a peak at $p=10^{-3}$ on both codes.\footnote{Longer chains do not remove it. On the $[[144,12,12]]$ BB code at $p=10^{-3}$, the long-start \obf{} running mean still sits at $4.99(2)$ against the reference $3.08(1)$ after $L=2.6\times10^{8}$ steps, about $130$ times the default length, falling by roughly $0.1$ per doubling, whereas \zh{} reaches $3.082(1)$ by $L=3.3\times10^{7}$.}

\zh{} agrees with the reference at every rate once the chains are long enough.
On the $[[108,8,10]]$ BB code at $p=10^{-2}$, the long-start entry differs from the reference by more than $3\sigma$ at the benchmark length, but a rerun with chains $20$ times longer removes the difference.\footnote{The rerun ($N=4\times10^{7}$, $K=48$) gives $+0.010(12)$ and $+0.014(12)$ from the long and short initializations, so the $+0.095(26)$ at $N=2\times10^{6}$ is a finite-chain residual that the convergence rule does not flag at that length. The rerun covers \zh{} only.}
On the $[[144,12,12]]$ BB code at $p=10^{-3}$, both entries agree with the reference but fail the convergence rule of \Cref{sec:bench1}, and only \zh{} reaches the $2\%$ target with chains about $16$ times longer than in this benchmark (\Cref{sec:bench2}).
Both deviations disappear with longer chains, so no large residual bias remains at the longer chain lengths tested.

Across \Cref{tab:meanw,tab:lowp}, the convergence rule of \Cref{sec:bench1} works well as a detector of bias, even though it does not see the reference at all.
Of the $16$ distinct bold entries, $15$ carry a dagger.
The one exception is the long-start \zh{} entry on the $[[108,8,10]]$ BB code at $p=10^{-2}$, whose bias is a finite-chain residual.
A stricter convergence threshold could flag this entry, and longer chains remove the deviation, as the rerun above shows.

\subsection{Time-to-accuracy benchmarks}

\begin{table*}[t]
\centering
\footnotesize
\setlength{\tabcolsep}{6pt}
\begin{tabular}{lccc}
\toprule
code & $p=10^{-1}$ & $p=10^{-3}$ & $p=10^{-5}$ \\
\midrule
surface $d=3$ & $0.010483(22)$ & $0.3821(27)$ & $1.10(6)\times10^{3}$ \\
surface $d=5$ & $0.009714(32)$ & $114.0(5)$ & $\geq 1.1\times10^{4}$ \\
surface $d=7$ & $0.00933(6)$ & $\geq 1.0\times10^{4}$ & $\geq 1.0\times10^{4}$ \\
toric $d=5$ & $0.005632(30)$ & $540(7)$ & $\geq 1.1\times10^{4}$ \\
toric $d=7$ & $0.004808(12)$ & $\geq 1.0\times10^{4}$ & $\geq 1.0\times10^{4}$ \\
concat.\ Steane, level $2$ & $0.008038(24)$ & $1.12(6)\times10^{3}$ & $\geq 9.1\times10^{3}$ \\
$\mathrm{HGP}(\text{Hamming})$ & $0.00375(31)$ & $0.3943(26)$ & $106(4)$ \\
$\mathrm{AT}(3,11)$ & $0.0045(9)$ & $0.4371(23)$ & $94(8)$ \\
$\mathrm{AT}(3,15)$ & $0.005242(18)$ & $0.268(23)$ & $37.7(33)$ \\
BB $[[72,12,6]]$ & $0.00398(33)$ & $4.992(27)$ & $\geq 2.1\times10^{3}$ \\
BB $[[144,12,12]]$ & $0.00197(21)$ & $0.217(20)$ & $\geq 2.2$ \\
BB $[[108,8,10]]$ & $0.00095(9)$ & $0.68(10)$ & $\geq 23$ \\
\bottomrule
\end{tabular}
\caption{Time-to-accuracy speedup of \zh{} over direct Monte Carlo, the $T_{2\%}$ of direct Monte Carlo divided by that of \zh{}, where $T_{2\%}$ (\Cref{eq:tta}) is the computation time after which the relative error stays below $2\%$. Parentheses give the bootstrap standard error defined in \Cref{sec:bench2}, and $\geq$ marks a lower bound, where direct Monte Carlo did not reach the $2\%$ target within its budget.}
\label{tab:dmcspeedup}
\end{table*}

\begin{table}[t]
\centering
\footnotesize
\setlength{\tabcolsep}{4pt}
\begin{tabular}{lccc}
\toprule
code & $p=10^{-1}$ & $p=10^{-3}$ & $p=10^{-5}$ \\
\midrule
surface $d=3$ & $0.4005(25)$ & $0.3336(30)$ & $0.3282(18)$ \\
surface $d=5$ & $0.79(22)$ & $0.2656(33)$ & $0.2661(22)$ \\
surface $d=7$ & $1.79(19)$ & $0.2464(28)$ & $0.36(4)$ \\
toric $d=5$ & $0.48(5)$ & $0.2602(26)$ & $0.2599(30)$ \\
toric $d=7$ & $1.12(6)$ & $0.2417(26)$ & $0.481(6)$ \\
concat.\ Steane, level $2$ & $0.38(4)$ & $0.4144(25)$ & $0.421(4)$ \\
$\mathrm{HGP}(\text{Hamming})$ & $0.53(10)$ & $0.224(24)$ & $0.353(4)$ \\
$\mathrm{AT}(3,11)$ & $0.74(18)$ & $\geq 590$ & $\geq 1.0\times10^{3}$ \\
$\mathrm{AT}(3,15)$ & $1.31(19)$ & $\geq 550$ & $\geq 530$ \\
BB $[[72,12,6]]$ & $0.51(7)$ & $21.1(10)$ & $\geq 1.4\times10^{3}$ \\
BB $[[144,12,12]]$ & $0.67(10)$ & $\geq 8.7$ & $\geq 12$ \\
BB $[[108,8,10]]$ & $0.58(7)$ & $\geq 1.6$ & $\geq 1.2$ \\
\bottomrule
\end{tabular}
\caption{Same speedups as \Cref{tab:dmcspeedup}, but against the one-bit-flip baseline (\obf). Parentheses give the bootstrap standard error defined in \Cref{sec:bench2}, and $\geq$ marks a lower bound, where \obf{} did not reach the $2\%$ target by the final checkpoint.}
\label{tab:obfspeedup}
\end{table}

The time-to-accuracy benchmarks measure the cost for each kernel to reach the $2\%$ target, the time to accuracy $T_{2\%}$ of \Cref{eq:tta}.
In \Cref{tab:dmcspeedup}, we compare the MCMC chains with direct Monte Carlo to find which is faster at each error rate on each code, and in \Cref{tab:obfspeedup}, we compare the two kernels to find which kernel is faster.

We first discuss the comparison with direct Monte Carlo.
At $p=10^{-1}$, direct Monte Carlo is faster for every code, by two to three orders of magnitude.
At $p=10^{-3}$, direct Monte Carlo is still faster on six of the twelve codes, though only by a factor of a few, whereas \zh{} is faster on the concatenated Steane code, the $[[72,12,6]]$ BB code, and the surface and toric codes at $d=5,7$.
At $p=10^{-5}$, \zh{} is faster on every code.
On the surface $d=3$ code, HGP(Hamming), and the two asymmetric toric codes, direct Monte Carlo reached the target, so the speedup is a measured value, and \zh{} is faster by factors of about $40$ to $10^{3}$.
On the concatenated Steane code, the three BB codes, and the surface and toric codes at $d=5,7$, direct Monte Carlo did not reach the target within its budget, which gives only a lower bound of the speedup.
These observations indicate that the crossover, the physical error rate below which \zh{} overtakes direct Monte Carlo, sits between $10^{-3}$ and $10^{-5}$ for six of the twelve codes.
For the other six, it sits above $10^{-3}$, and the speedup keeps growing below $10^{-3}$.
The increase in speedup as the physical error rate $p$ decreases is explained by the direct Monte Carlo cost, which grows as $1/P_L$, while the MCMC cost changes little.

We now turn to the comparison between the two kernels, \zh{} and \obf{}.
On the seven codes on which \obf{} is not trapped, namely the surface codes at $d=3,5,7$, the toric codes at $d=5,7$, the concatenated Steane code, and the HGP(Hamming) code, \zh{} is slower than \obf{} at $p=10^{-3}$ and $10^{-5}$ by a factor between $2$ and $5$, which can be explained by the constant per-step overhead of the locally balanced proposal.
Where \obf{} is accurate to within the target and fast, it is therefore the cheaper kernel.
At $p=10^{-1}$, where we found that both kernels reach the $2\%$ target within the first few checkpoints, \zh{} has no advantage either.\footnote{Since consecutive checkpoints differ by a factor of two in $L$, each $T_{2\%}$ is resolved only to within a factor of two, so the entries of order one in this column do not imply the superiority of \zh{}.}

On the five trapping codes, namely the asymmetric toric codes $\mathrm{AT}(3,11)$ and $\mathrm{AT}(3,15)$ and the BB codes $[[72,12,6]]$, $[[144,12,12]]$, and $[[108,8,10]]$, by contrast, \zh{} is faster at $p=10^{-3}$ and $10^{-5}$.
On the two asymmetric toric codes and on the $[[72,12,6]]$ BB code, \zh{} reaches the target sooner, by a factor of $21$ where \obf{} reaches the target and by lower bounds of $530$ to $1.4\times10^{3}$ elsewhere.
On the $[[144,12,12]]$ and $[[108,8,10]]$ BB codes, \obf{} did not reach it within its budget, while \zh{} did, giving only a lower bound on the speedup.

On the $[[72,12,6]]$ BB code, \obf{} shows no bias beyond $3\sigma$ at the end of the accuracy benchmark but is slow to reach the target, as the daggers on its \obf{} entries in \Cref{tab:meanw} show.
This observation indicates that the trapping can increase the time to accuracy without a detectable bias on that code, and that the speedup of \zh{} comes mainly from mitigating the trapping of \obf{} rather than from faster mixing within a sector.

Both benchmarks in this section concern the mean fault weight, and \Cref{sec:plcost} discusses what they imply for the cost of estimating the logical error rate, the quantity that matters in practice.

\section{Discussion}
\label{sec:discussion}

\subsection{How to verify a new code}
\label{sec:recipe}

The results above give the following recipe for verifying a previously unchecked code.
An MCMC-independent estimate, by the weight-stratified reference (\Cref{sec:setup-stratified}) or by direct Monte Carlo (\Cref{sec:decoding}), is preferable to MCMC wherever it is affordable, since it cannot be trapped.
The recipe addresses the situation in which the MCMC-independent estimate is affordable at the highest error rates of the ladder but not at the lowest, so that MCMC is a candidate for the remainder.

It has four steps.
\begin{enumerate}
\item Choose the kernel.
We recommend \zh{} as the default, with exchange size $m=2$ and cluster-exchange probability $p_{\mathrm{cluster}}=0.3$, the values used in the main benchmarks.
In \Cref{app:mexch,app:pcluster}, we sweep both parameters at $p=10^{-4}$ under the protocol of \Cref{sec:bench2}.
Near these defaults, the speed depends little on either parameter on most codes.
\obf{} is faster by a constant factor on codes without observed trapping (\Cref{tab:obfspeedup}), but we recommend it only where its ergodicity is proven~\cite{bravyi2013simulation}.
Where no such proof exists, a broad basis-weight spread (\Cref{eq:basis-weight-spread}) marks \obf{} as risky (\Cref{sec:results}).

\item Check that the Markov chains reproduce the MCMC-independent reference at every error rate it reaches.
Agreement there is necessary but not sufficient, since it says nothing about the lower error rates the reference does not reach.
For a quick check, use the mean fault weight $\mathbb{E}_{\mathcal{F}}[w]$, which is the simplest statistic of the fault manifold and can be calculated at no extra cost.
For a stronger check, run the splitting method down to the lowest error rate the reference reaches and compare the logical error rate $P_L$ itself.

\item Below the reach of the MCMC-independent reference, apply all four parts of the convergence rule of \Cref{sec:bench1}, which uses only the statistics of the chains themselves.
Part (iii), the agreement of the estimates from chains started with deliberately different states, is the decisive one.
It is the only part that can catch chains trapped in one region of the fault manifold, where every statistic of the chains looks healthy.
The initial states should be as far apart as the code allows, for example faults from different logical sectors.
A convenient pair is the maximum- and minimum-weight logical-$X$ basis elements.
However, passing the rule does not prove correctness, since it detects only failures that show up in the statistics of the chains.
Other statistics, such as the estimated integrated autocorrelation time $\iat$ or the Gelman--Rubin statistic $\hat R$~\cite{gelman1992inference,vehtari2021rank}, may be worth checking as well.

\item When a warning appears, first increase the chain budget and repeat steps 2 and 3.
If the warning persists, try a more sophisticated kernel.
\end{enumerate}

The recipe defines what \emph{fast} and \emph{accurate} mean for a rare-event sampler.
Fast means that, below the crossover of \Cref{tab:dmcspeedup}, the chains reach a given accuracy sooner than direct Monte Carlo, and that the speedup grows as the logical error rate $P_L$ falls.
Compared with \obf{}, the chains are faster on the trapping codes (\Cref{tab:obfspeedup}).
Accurate means agreement with the MCMC-independent reference where it is available, and an estimate that does not depend on the initial state and passes the convergence rule under the reported budget.
The latter is particularly important below the reach of the MCMC-independent reference.

\subsection{From the mean weight to the cost of estimating the logical error rate}
\label{sec:plcost}

All the benchmarks so far concern the mean fault weight $\mathbb{E}_{\mathcal{F}}[w]$, whereas the quantity of interest is the logical error rate $P_L$, which is evaluated by the splitting method.
A faster estimate of the mean weight does not by itself imply a faster estimate of the logical error rate, since the two estimators spend their computation time differently.
In the following, we break down the cost of each in turn.

\begin{figure*}[t]
  \centering
  \includegraphics[width=\textwidth]{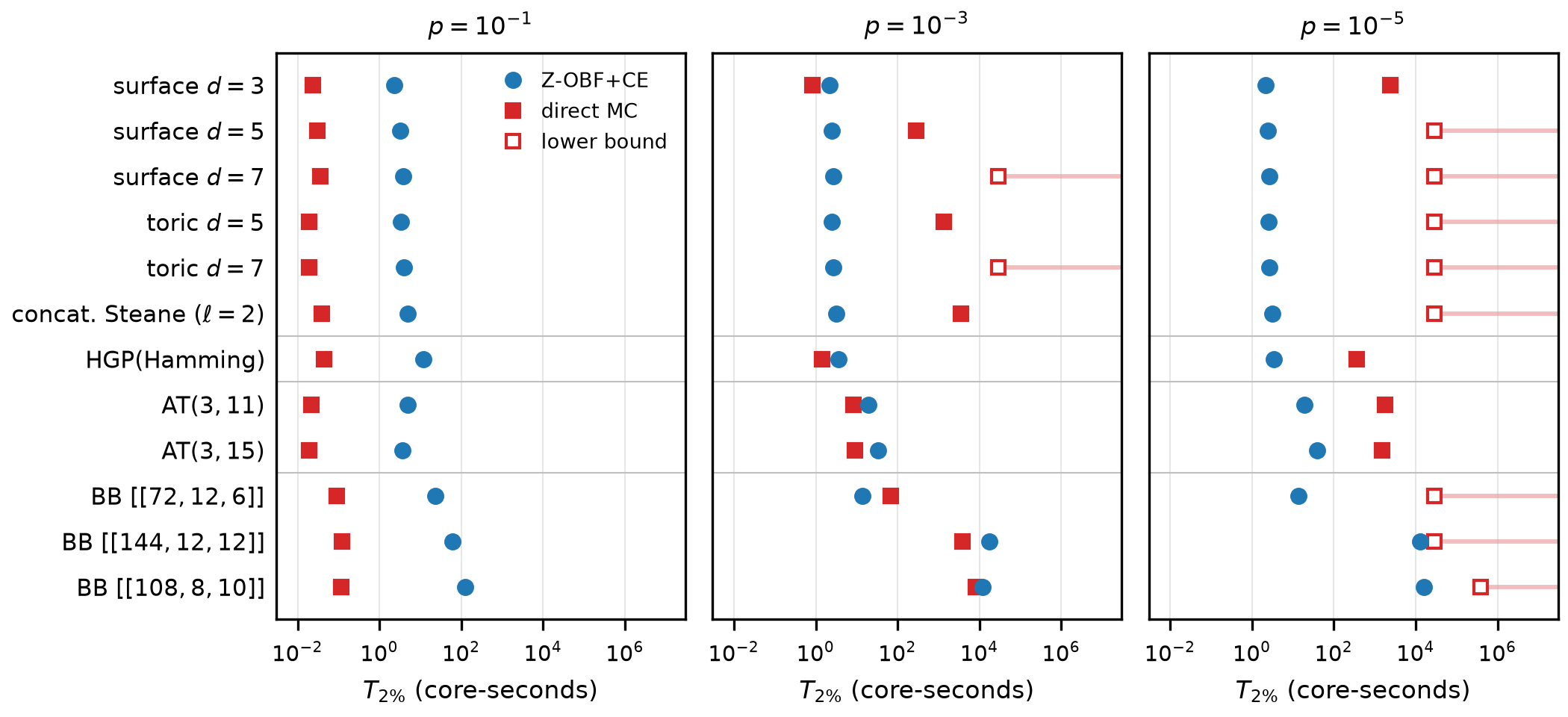}
  \caption{Time to accuracy $T_{2\%}$ of \Cref{eq:tta} in core-seconds for \zh{} (circles) and direct Monte Carlo (squares) on the twelve benchmark codes.
Each panel corresponds to one physical error rate $p$ and each row to one code.
An open square with a line extending to the right marks a lower bound, where direct Monte Carlo did not reach the $2\%$ target within its budget, as in \Cref{tab:dmcspeedup}.
Each symbol is the mean over the ten runs of \Cref{sec:bench2}, and its horizontal error bar gives the bootstrap standard error of that mean over the runs.
Most error bars are smaller than the symbols.
}
  \label{fig:ttawalltime}
\end{figure*}

In \Cref{fig:ttawalltime}, we plot, in core-seconds, the times to accuracy whose ratios give the speedups in \Cref{tab:dmcspeedup}.
Direct Monte Carlo obtains the logical error rate and the mean weight from the same run.
The fraction of draws that are faulty is the logical error rate, and their average weight is the mean weight.
A relative error of $\epsilon=2\%$ on the logical error rate requires about $1/\epsilon^2=2500$ faulty draws (\Cref{sec:decoding}).

On the other hand, the mean weight needs fewer faulty draws.
Its relative error after $N_{\mathrm{faulty}}$ faulty draws is, by \Cref{eq:dmc-se},
\begin{align}  
  \frac{1}{\mathbb{E}_{\mathcal{F}}[w]} \sqrt{\frac{\mathrm{Var}_{\mathrm{ideal}}}{N_{\mathrm{faulty}}}}\text{,}
\end{align}
with $\mathrm{Var}_{\mathrm{ideal}}$ the variance of the weight over the fault manifold.
Therefore, a relative error of $\epsilon$ needs $\mathrm{Var}_{\mathrm{ideal}}/(\mathbb{E}_{\mathcal{F}}[w]^2\epsilon^2)$ faulty draws.
If the physical error rate is very small, the ratio $\mathrm{Var}_{\mathrm{ideal}}/\mathbb{E}_{\mathcal{F}}[w]^2$ becomes small because the conditional fault distribution concentrates on errors whose weight is close to the minimum fault weight $w_{\min}\coloneqq\min_{e\in\mathcal{F}}w(e)$.
The time of direct Monte Carlo in the figure is therefore a lower bound on its cost of estimating the logical error rate.

The splitting method instead spends its time in three stages.
The first estimates the logical error rate at the top of the ladder, $P_L(p_0)$, by an unbiased method such as direct Monte Carlo.
The second runs the Markov chains at each physical error rate $p_\ell$ of the ladder, the decreasing sequence of \Cref{sec:splitting}.
The third combines the samples at adjacent physical error rates into the ratios $R_\ell$ by the BAR method of \Cref{eq:bar}.
The total computation time is therefore given by
\begin{align}
  T_{\mathrm{split}} = T_{\mathrm{top}} + \sum_{\ell=0}^{J} T_{\mathrm{MCMC}}(p_\ell) + T_{\mathrm{BAR}}.
  \label{eq:split-cost}
\end{align}
The first term is small, because faulty draws are frequent at the top of the ladder.
The last term is small as well, because the BAR method only solves the scalar self-consistent equation of \Cref{eq:bar} on the stored samples.
The Markov chains in the second stage therefore dominate the total.

We now estimate this dominant term on the benchmark codes.
We assume that estimating a ratio $R_\ell$ costs as much as estimating the mean weight at the same physical error rate.
We further assume that the levels we did not benchmark cost no more than the most expensive level we did.
Under these assumptions, the chain cost at every level is estimated to be at most the largest $T_{2\%}$ of \zh{} in \Cref{fig:ttawalltime}. 
Then, the total time is at most this time multiplied by the number of physical error rates in the ladder.
For a ladder whose adjacent physical error rates have a fixed ratio, that number scales as $\log(p_0/p_J)$.
The direct Monte Carlo cost, by contrast, scales as the inverse of the logical error rate.

In \Cref{tab:plcost}, we evaluate this estimate for a $2\%$ target on the logical error rate at $p=10^{-5}$, with a ladder of $21$ physical error rates starting from $10^{-1}$.
The relative errors of the $20$ ratios add in quadrature, and hence reaching $2\%$ at $p=10^{-5}$ costs $20$ times the computation time at every level.
Under these assumptions, the splitting method is cheaper than direct Monte Carlo on all twelve codes, by a factor of about $10$ or more.
On eight of the twelve codes, direct Monte Carlo collected no faulty draw within its budget, so its time, and hence the factor, is a lower bound.

\begin{table*}[t]
\centering
\footnotesize
\setlength{\tabcolsep}{6pt}
\begin{tabular}{lccc}
\toprule
code & splitting & direct MC & ratio \\
\midrule
surface $d=3$ & $960$ & $6.0\times10^{6}$ & $6.2\times10^{3}$ \\
surface $d=5$ & $1.3\times10^{3}$ & $\geq 7.2\times10^{7}$ & $\geq 5.3\times10^{4}$ \\
surface $d=7$ & $1.6\times10^{3}$ & $\geq 7.2\times10^{7}$ & $\geq 4.5\times10^{4}$ \\
toric $d=5$ & $1.4\times10^{3}$ & $\geq 7.2\times10^{7}$ & $\geq 5.1\times10^{4}$ \\
toric $d=7$ & $1.7\times10^{3}$ & $\geq 7.2\times10^{7}$ & $\geq 4.2\times10^{4}$ \\
concat.\ Steane, level $2$ & $2.1\times10^{3}$ & $\geq 7.2\times10^{7}$ & $\geq 3.5\times10^{4}$ \\
$\mathrm{HGP}(\text{Hamming})$ & $5.0\times10^{3}$ & $9.1\times10^{5}$ & $180$ \\
$\mathrm{AT}(3,11)$ & $8.2\times10^{3}$ & $4.5\times10^{6}$ & $560$ \\
$\mathrm{AT}(3,15)$ & $1.7\times10^{4}$ & $3.7\times10^{6}$ & $220$ \\
BB $[[72,12,6]]$ & $9.5\times10^{3}$ & $\geq 7.2\times10^{7}$ & $\geq 7.5\times10^{3}$ \\
BB $[[144,12,12]]$ & $7.4\times10^{6}$ & $\geq 7.2\times10^{7}$ & $\geq 9.7$ \\
BB $[[108,8,10]]$ & $6.8\times10^{6}$ & $\geq 7.2\times10^{7}$ & $\geq 10$ \\
\bottomrule
\end{tabular}
\caption{Cost of estimating the logical error rate to $2\%$ relative accuracy at $p=10^{-5}$ (\Cref{sec:plcost}) in core-seconds. The splitting entry is the estimate under the assumptions of \Cref{sec:plcost} for a ladder of $21$ physical error rates, the largest \zh{} time of \Cref{fig:ttawalltime} multiplied by $21\times20$. The direct Monte Carlo (direct MC) entry is its time for $2500$ faulty draws, and the inequality sign $\geq$ marks a lower bound, where direct Monte Carlo collected no faulty draw within its budget. The last column gives the direct Monte Carlo cost divided by the splitting cost.}
\label{tab:plcost}
\end{table*}

Between \obf{} and \zh{}, the estimate of $P_L(p_0)$, the ladder, and the BAR step are shared, and only the cost of the chains at each level differs.
The speedups of \Cref{tab:obfspeedup} are therefore expected to carry over to the cost of estimating the logical error rate under the same assumptions.

\subsection{Scope of our method}
\label{sec:scope}

The results above were obtained on specific codes, under bit-flip noise, and with the decoders of \Cref{sec:setup-codes}.
Carrying our method to another code, noise model, or decoder requires three things.

First, the kernel of the Markov chains, \zh{} in this paper, has to reach every configuration of the fault manifold of the code.
A proof of ergodicity guarantees this, and without one, a comparison against an MCMC-independent reference provides an empirical check.

Second, the decoder has to define the same fault manifold at every level of the ladder.
This is a requirement of the splitting method rather than of the sampler.

Third, the cluster exchange has to keep an acceptance probability of order one, $O(p^0)$, under the noise model, that is, one not suppressed by a power of the physical error rate $p$.
The locally balanced proposal can be extended straightforwardly wherever \obf{} can, since it only reweights the single-bit flips of \obf{}, while the cluster exchange is accepted with probability one only where the target distribution $\pi_p$ depends on the configuration through its Hamming weight alone, provided that the proposed configuration remains faulty.
We discuss these three conditions in turn.

As for the first condition, ergodicity, neither kernel is proven ergodic on general codes.
Cluster exchange connects configurations of equal weight directly, and single flips connect configurations whose weight differs by one.
Note that the locally balanced proposal only reweights the single-bit flips of \obf{}, and therefore has the same ergodicity properties as \obf{}.
The only proof of ergodicity we know of covers \obf{} on the surface code in the geometries of Ref.~\cite{bravyi2013simulation}.

As for the second condition, the level ratios of \Cref{eq:bar} assume that the fault manifold is the same at two adjacent levels.
However, a decoder that takes the physical error rate as an input, such as \bposd{}, can have a different fault manifold along the ladder.
Because the fault manifold is dominated by its low-weight configurations at low physical error rates, the splitting method is expected to remain approximately valid, as long as the decoder fails on the same low-weight configurations at adjacent levels.
This approximation requires a separate check for \bposd{} on the BB codes, whose failure fractions vary substantially with the error rate (\Cref{sec:setup-stratified}).
If one wants to obtain an unbiased estimate of the logical error rate with different decoders at different physical error rates, one needs to extend the splitting method itself.

As for the third condition, it concerns the noise model rather than the sampler.
Both the trapping of \obf{} and its removal by \zh{} are observed in this paper under bit-flip noise.
The noise models of practical interest are more general.
They include depolarizing noise on the data qubits, phenomenological noise with faulty syndrome measurements, circuit-level noise with errors at every location of the syndrome-extraction circuit, and biased or correlated variants of each.
A noise model enters the kernels through the configuration space and through the target distribution $\pi_p$.

Both the trapping and its removal come from the target distribution $\pi_p$.
The trapping arises because the target distribution $\pi_p$ suppresses the high-weight configurations that separate the logical sectors, so a single flip, which has to pass through them, is rarely accepted (\Cref{sec:baseline}).
Cluster exchange avoids them because it moves between configurations of the same weight.
Such a move is accepted with probability one when the target distribution $\pi_p$ depends on the configuration through its Hamming weight alone (\Cref{eq:cluster_accept}), which is why it removes the trapping under bit-flip noise.

Let us consider in which noise models our kernel remains effective.
The first example is depolarizing noise, where each qubit carries one of four Pauli states instead of two.
The target distribution $\pi_p$ depends on the configuration only through its support weight, the number of qubits carrying a nonidentity Pauli error, so the cluster exchange is still accepted with probability one.
We test the validity of our kernel in \Cref{app:depol} for the $[[144,12,12]]$ BB code.
There, we find that \obf{} is again biased from both the short and the long initializations at $p=10^{-3}$ and $10^{-4}$ and fails the convergence rule, while \zh{} agrees with the MCMC-independent reference from both starts at every error rate and passes the rule.
The trapping of \obf{} and its reduction by \zh{} are therefore observed under depolarizing noise as well.

This argument holds for the biased Pauli noise model, where each qubit still suffers an error independently but the Pauli errors $X$, $Y$, and $Z$ occur with different probabilities.
There, the target distribution $\pi_p$ depends on the Pauli types of the errors as well as on their number, so a cluster exchange that changes the Pauli type of an error has an acceptance ratio equal to the ratio of the probabilities of the two error types.
This ratio is still of order one, $O(p^0)$, if every type scales with the physical error rate $p$ in the same way.
In that case, the acceptance of the cluster exchange is not suppressed as $p$ decreases, and we expect our kernel to remain effective in the biased Pauli noise model as well.

The next one is phenomenological noise, where the syndrome measurements are faulty as well and we need several rounds of syndrome measurement.
An error location is now a data qubit or a syndrome measurement at a given round.
Even in this noise model, the cluster exchange can be defined as a weight-preserving move that relocates an error from one location to another.
When the move relocates an error between a data qubit and a measurement, the acceptance ratio is $p_{\mathrm{m}}(1-p_d)/(p_d(1-p_{\mathrm{m}}))$ or its inverse for data and measurement error probabilities $p_d$ and $p_{\mathrm{m}}$.
It is of order one, $O(p^0)$, as long as $p_d$ and $p_{\mathrm{m}}$ scale in the same way with the physical error rate.
Under this condition, our kernel is expected to work well.

The last one is circuit-level noise, where an error location is now a gate, a measurement, or an idle qubit of the syndrome-extraction circuit at a given time step, with its own set of Pauli errors.
A natural extension of the cluster exchange is a move that relocates errors while preserving their number.
Its acceptance ratio is the ratio of the error odds of the locations involved, which is of order one for the same reason as above.
In the SI1000 model of Ref.~\cite{gidney2021honeycomb}, for example, every location has an error probability that is a fixed multiple of $p$, from $p/10$ for single-qubit gates to $5p$ for measurement flips.
Therefore, the acceptance ratio of a relocation is independent of $p$ but can be as small as $1/50$.
Whether the cluster-exchange-like move still mixes well under such a model has to be checked in practice.

\section{Conclusion and outlook}
\label{sec:conclusion}

In this work, we have studied which sampling methods are suitable for rare-event simulation of general quantum codes.
Direct Monte Carlo is unbiased but prohibitively slow at low physical error rates, while Markov chain Monte Carlo (MCMC) with the one-bit-flip (\obf{}) kernel is faster but biased on some codes.
We then proposed a new MCMC method combining the cluster exchange and locally balanced proposals, which aims to remove the bias of the one-bit-flip kernel.

The locally balanced proposal of \Cref{sec:zanella} (\zk{}) flips one bit, chosen with a weight that favors the flips likely to be accepted under the target distribution $\pi_p$, instead of a uniformly random bit.
The cluster exchange of \Cref{sec:cluster} (\ck{}), which moves errors between qubits without changing their number, is designed to connect the logical sectors without a weight-dependent acceptance penalty.

We performed two benchmark tests on twelve codes, the accuracy benchmark and the time-to-accuracy benchmark, with our kernel \zh{}, the one-bit-flip kernel \obf{}, and direct Monte Carlo.
The accuracy benchmark compares the estimate of the mean fault weight at a fixed budget with an MCMC-independent reference.
The \zh{} kernel reproduces the reference on all codes, while the one-bit-flip kernel is biased on the asymmetric toric codes and on the $[[144,12,12]]$ and $[[108,8,10]]$ BB codes.
The time-to-accuracy benchmark measures the computation time until the estimate of the mean fault weight comes within $2\%$ of the reference and stays there.
At the lowest error rate, the \zh{} kernel reaches the target accuracy faster than direct Monte Carlo on all codes, by more than four orders of magnitude on some of them.
Against the one-bit-flip kernel, the \zh{} kernel is faster on the five codes where the one-bit-flip kernel is trapped.
On the asymmetric toric codes and the $[[72,12,6]]$ BB code, the speedup is $21$ where the one-bit-flip kernel reaches the target and at least $530$ to $1.4\times10^{3}$ elsewhere.
Where the one-bit-flip kernel is not trapped, the \zh{} kernel is slower by a factor between $2$ and $5$.

We turn these benchmarks into a recipe for verifying a new code in \Cref{sec:recipe}.
We then estimate in \Cref{sec:plcost} how the speed advantage of our kernel carries over to the cost of estimating the logical error rate under stated assumptions, since the chains dominate the cost of the splitting method.
In \Cref{sec:scope}, we finally discuss the three conditions for applying the method to other codes, decoders, and noise models.
While we assumed code-capacity bit-flip noise (and depolarizing noise) throughout this work, our method is expected to extend to other noise models such as biased code-capacity noise or phenomenological noise as long as the acceptance ratio of the cluster exchange stays of order one, $O(p^0)$, as $p$ decreases.

Three questions remain open.
The first is the ergodicity of the kernel on general codes.
The only proof we know of covers the one-bit-flip kernel on the surface code in the geometries of Ref.~\cite{bravyi2013simulation}, and the evidence for our kernel is empirical.
The second is the extension of the splitting method itself to a decoder whose fault manifold changes with the physical error rate, such as \bposd{}, since the level ratios of the ladder assume the same fault manifold at adjacent levels.
The third is the effectiveness of the kernel under circuit-level noise.
Although the cluster exchange can be extended to a move that relocates errors while preserving their number, it is an open question whether this kind of move can remove the bias under such a noise model.

Finally, fault-tolerant quantum computation aims at logical error rates of order $10^{-12}$ or below~\cite{beverland2025failfast}.
At such rates, direct sampling needs more than $10^{12}$ decoding trials for every failure it observes, far beyond any practical budget.
The MCMC method offers a solution to this problem, since it generates samples that follow the conditional fault distribution $\pi_{p\mid\mathcal{F}}$ with a Markov chain.
This requires a kernel that explores the fault manifold without bias and within a reasonable computational budget.
We hope that this work is a step toward that goal.

\section*{Acknowledgments}

We thank Michael E.\ Beverland for valuable discussions at the QEC 2026 conference.

\section*{Use of AI tools}

Anthropic models (Claude Sonnet 4.6, Opus 4.7, 4.8, 5, and 5.5, and Fable 5 and 5.1) and an OpenAI model (GPT-6 Astra) were used in this work.
They assisted with drafting the manuscript and with writing the simulation and analysis code.
The authors directed the computations, reviewed the generated code and text, and checked the reported numbers against the simulation outputs.
All figures were generated by the authors' plotting scripts from the simulation outputs.
The authors take responsibility for the final code, analysis, and manuscript.

\section*{Funding}

J.H.\ is supported by the MEXT Quantum Leap Flagship Program (MEXT Q-LEAP) Grant No.~JPMXS0120319794, JST COI-NEXT Grant No.~JPMJPF2014, JST Moonshot R\&D Grant No.~JPMJMS256E, and JST CREST Grant No.~JPMJCR24I3.
K.M.\ is supported by JSPS KAKENHI Grant Nos.~JP22H04917 and JP25K17384.

\section*{Data availability}

The code, the cluster job scripts, the simulation outputs, and the reference estimates that generate every table and figure in this paper are available on Zenodo at \url{https://doi.org/10.5281/zenodo.23660599}.
A standalone Python implementation of the exact enumeration, direct Monte Carlo, and BAR splitting estimators, including the kernels used here, is available as the package \texttt{mcmc4qec} at \url{https://github.com/junichiharuna/mcmc4qec}.

\appendix
\numberwithin{equation}{section}

\section{Depolarizing noise}
\label[appendix]{app:depol}

In this appendix, we extend the benchmark of \Cref{sec:results} from bit-flip noise to the joint $X/Z$ fault manifold under depolarizing noise.
We thereby test whether the trapping and its removal are properties of the sampler rather than artifacts of the error channel.

\subsection{The joint target and fault manifold}
\label{app:depol-target}
Under depolarizing noise, each qubit is independently left alone with probability $1-p$ or hit by one of the Pauli errors $X,Y,Z$ with probability $p/3$ each.
We represent the resulting Pauli error by its $X$- and $Z$-parts $(e_X, e_Z)\in\GF^n\times\GF^n$, and a qubit carries a $Y$ error when it appears in both parts.
The configuration probability factorizes over qubits and depends only on the \emph{support weight} $w = |\,e_X \vee e_Z\,|$, the number of qubits with a nonidentity Pauli error, and is given by
\begin{align}
  \pi_p(e_X, e_Z) = \left(\frac{p}{3}\right)^{\!w}(1-p)^{\,n-w}\text{.}
  \label{eq:pi-depol-joint}
\end{align}
Restricting and normalizing this distribution on the joint fault manifold defined below gives the sampling target.
In this appendix, $w$ denotes the support weight rather than the Hamming weight.

The $X$- and $Z$-errors are decoded separately, the $X$-errors from the $Z$-syndrome and the $Z$-errors from the $X$-syndrome, which gives the residuals $r_X = e_X \oplus \mathcal{D}_X(H_Z e_X)$ and $r_Z = e_Z \oplus \mathcal{D}_Z(H_X e_Z)$.
A configuration is faulty when either residual fails its logical test,
\begin{align}
  (e_X,e_Z)\in\mathcal{F}_{\mathrm{dep}} \iff L_Z\, r_X \neq 0 \ \ \text{or}\ \ L_X\, r_Z \neq 0 .
  \label{eq:faulty-depol}
\end{align}

\subsection{The local and cluster moves}
\label{app:depol-moves}
In this paper, we generalize the one-bit flip to depolarizing noise as follows.
A single move multiplies one qubit by $X$ or by $Z$, which flips the bit $(e_X)_i$ or $(e_Z)_i$.
A qubit in the state $I$ thus moves to $X$ or $Z$, a qubit in the state $X$ or $Z$ moves to $I$ or $Y$, and a qubit in the state $Y$ moves to $Z$ or $X$.
Multiplication by $Y$ could also be included as a single move, but here it is reached by two successive moves.

The locally balanced proposal of \Cref{sec:zanella} applies to these moves.
Each of the $2n$ candidate moves changes the support weight by $\Delta w\in\{-1,0,+1\}$, and thus multiplies the probability of \Cref{eq:pi-depol-joint} by $\{p/[3(1-p)]\}^{\Delta w}$.
We draw one of the candidate moves with probability proportional to the square-root balancing weight of \Cref{eq:zanella_weight}, which is the square root of this ratio as in the bit-flip case.
A move therefore carries balancing weight $\sqrt{p/[3(1-p)]}$ when it raises the support weight ($I\to X$, $I\to Z$), $\sqrt{3(1-p)/p}$ when it lowers it ($X\to I$, $Z\to I$), and $1$ when it leaves it unchanged ($X\!\leftrightarrow\! Y$, $Z\!\leftrightarrow\! Y$).

Each qubit in the state $I$ has two candidate moves, both of which raise the support weight.
Each qubit in the state $X$ or $Z$ has one candidate move that lowers the support weight and one that leaves it unchanged.
Each qubit in the state $Y$ has two candidate moves, both of which leave the support weight unchanged, so the proposal normalizer is given by
\begin{align}
  \Lambda(e) &= 2\sqrt{\frac{p}{3(1-p)}}\, n_I + 2\, n_Y \notag\\
  &\quad {}+ \left(\sqrt{\frac{3(1-p)}{p}}+1\right)(n_X+n_Z),
  \label{eq:zanella-depol-Z}
\end{align}
with $n_I,n_X,n_Z,n_Y$ the counts of qubits in each Pauli state.
The acceptance is again $A=\min(1,\,\Lambda(e)/\Lambda(e'))$ as in \Cref{eq:zanella_accept}.

The weight-preserving cluster exchange generalizes to the joint manifold as a move that preserves the support weight, with the same exchange size $m$ as in \Cref{sec:cluster}.
Each step picks one of the following two submoves with equal probability.
One is relocation, which clears $m$ occupied qubits and fills $m$ empty ones.
The other is relabeling, which replaces the Paulis on $m$ occupied qubits.
In both submoves, the new Paulis are drawn uniformly from $X,Y,Z$.
Both submoves keep the support weight $w=|e_X\vee e_Z|$ fixed and are symmetric, and because the target distribution $\pi_p$ depends only on $w$ (\Cref{eq:pi-depol-joint}), the Metropolis--Hastings ratio of \Cref{eq:cluster_accept} is again one.
The proposal is therefore accepted whenever the candidate is still faulty, exactly as in the bit-flip case.

\begin{figure*}[t]
  \centering
  \includegraphics[width=0.85\textwidth]{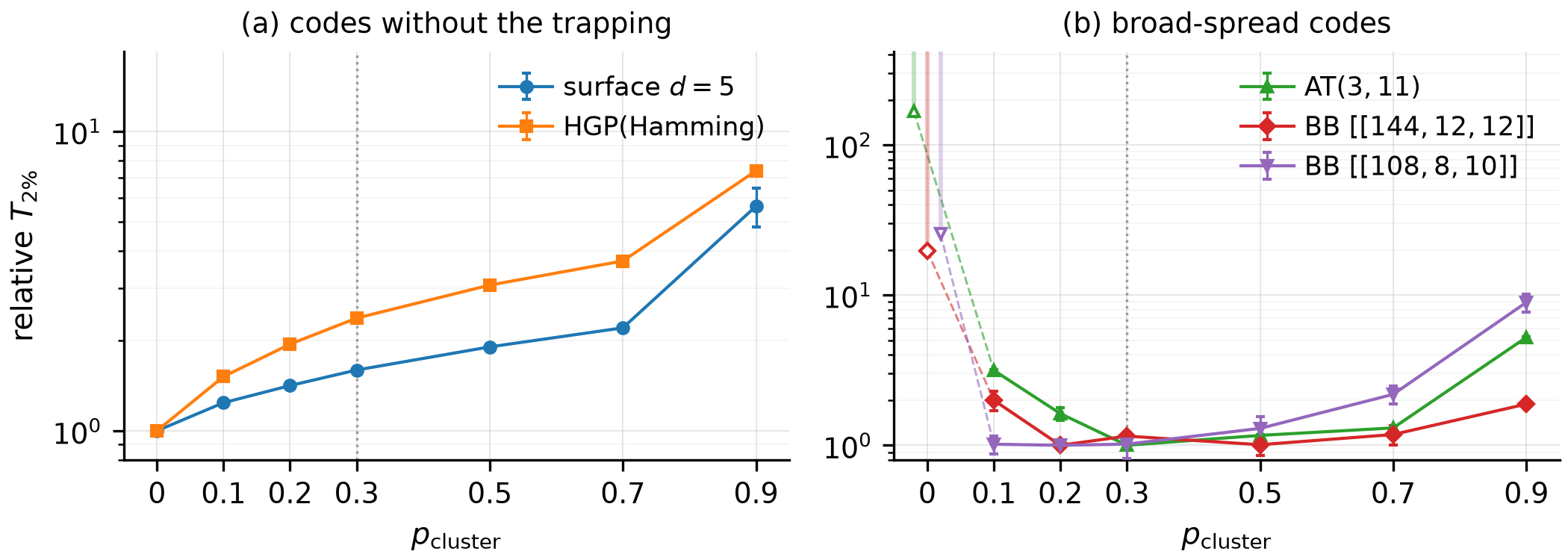}
  \caption{Time to accuracy $T_{2\%}$ of \Cref{eq:tta} versus the cluster-exchange probability $p_{\mathrm{cluster}}$ at the exchange size $m=2$ and the error rate $p=10^{-4}$, measured as in \Cref{tab:dmcspeedup} over five runs of $50$ chains.
Panel (a) shows the two codes without the trapping and panel (b) the three broad-spread codes.
Each curve shows the time to accuracy divided by the smallest reached value for that code, so the fastest reached setting of each code has the value $1$.
Error bars give the standard error of this ratio, propagated from the bootstrap standard errors of the two times.
An open marker with a line upward, joined to its curve by a dashed line, is a lower bound where at least one run did not reach the target within $L=10^{3}\cdot2^{11}\approx2.0\times10^{6}$ steps.
The three lower bounds in panel (b) all lie at $p_{\mathrm{cluster}}=0$ and are shifted horizontally for visibility.
The dotted line marks the probability $p_{\mathrm{cluster}}=0.3$ used in the main text.}
  \label{fig:pcluster}
\end{figure*}

\subsection{The benchmark}
\label{app:depol-comparison}
We apply the first benchmark of \Cref{sec:bench1} to the broad-spread $[[144,12,12]]$ BB code, on which the baseline is trapped under bit-flip noise, at the error rates $p=10^{-2}$, $10^{-3}$, and $10^{-4}$ of \Cref{tab:lowp}.
The setup is that of \Cref{sec:bench1}, namely $50$ chains from each of the short and long initializations, the $5\iat$ burn-in, and the four-part convergence rule, except that each chain has $N=2\times10^{7}$ steps, ten times the benchmark length.

The reference is the weight-stratified estimate of \Cref{sec:setup-stratified}, which carries over to the joint fault manifold of \Cref{eq:faulty-depol} with the support weight in place of the Hamming weight.
The support weight is binomially distributed and the Pauli types on the support are i.i.d.\ uniform, so \Cref{eq:stratified,eq:stratified-ew} hold with the failure fraction $f_w\coloneqq\Pr\bigl((e_X,e_Z)\in\mathcal{F}_{\mathrm{dep}}\mid w\bigr)$ of the joint weight class.
We measure $f_w$ as in \Cref{sec:setup-stratified}, by enumeration or uniform sampling over the $\binom{n}{w}3^{w}$ patterns of a class, separately for each error rate $p$ supplied to the \bposd{} decoder, and with the truncation $w_{\max}=12$.
Against this reference, we compare the one-bit-flip baseline and the recommended \zh{} kernel at its default parameters (\Cref{tab:depol}).

\begin{table}[t]
\centering
\begin{tabular}{lcccc}
\toprule
$p$ & reference $\mathbb{E}_{\mathcal{F}}[w]$ & init & \obf{} & \zh{} \\
\midrule

$10^{-2}$ & $4.314(23)$ & short & $+0.05(10)$ & $+0.010(24)$ \\
          &              & long  & $+0.22(9)$ & $+0.007(23)$ \\
$10^{-3}$ & $3.097(5)$ & short & $\mathbf{+1.27(11)}^\dagger$ & $0.000(6)$ \\
          &              & long  & $\mathbf{+1.15(14)}^\dagger$ & $0.000(6)$ \\
$10^{-4}$ & $3.0487(18)$ & short & $\mathbf{+0.46(7)}^\dagger$ & $+0.0007(18)$ \\
          &              & long  & $\mathbf{+0.42(6)}^\dagger$ & $+0.0006(18)$ \\
\bottomrule
\end{tabular}
\caption{Accuracy benchmark under depolarizing noise on the $[[144,12,12]]$ BB code under the setup of \Cref{sec:bench1}, with chains of $2\times10^{7}$ steps, ten times the benchmark length.
The entries are the estimated signed bias $\hat{\mathbb{E}}_{\mathcal{F}}[w]-\mathbb{E}_{\mathrm{ref}}$ of the joint support weight $w=|e_X\vee e_Z|$ against the weight-stratified reference, from the minimum-weight (short) and the maximum-weight (long) initialization.
Parentheses give standard errors, and daggers and bold follow the convention of \Cref{tab:meanw}.}
\label{tab:depol}
\end{table}

On this code, the verdict is the same as under bit-flip noise.
\zh{} agrees with the reference from both starts at every error rate, within $0.4\sigma$, and all six of its entries pass the convergence rule.
At $p=10^{-3}$ and $10^{-4}$, \obf{} overestimates the support weight by $0.4$ to $1.3$ from either initialization, more than six standard errors from the reference, and all four of these entries fail the convergence rule, consistent with trapping from both starts.
At $p=10^{-2}$, the two \obf{} entries pass the convergence rule and lie within $2.3\sigma$ of the reference.
The bit-flip picture of \Cref{tab:lowp} thus persists on this code under depolarizing noise, down to $p=10^{-4}$.

\section{Parameter sweeps of Z-OBF+CE}

In this appendix, we record the two parameter sweeps of our kernel.
We examine the effect of the exchange size $m$ in Sec.~\ref{app:mexch} and the sensitivity to the probability $p_{\mathrm{cluster}}$ of proposing a cluster exchange in Sec.~\ref{app:pcluster}.

\begin{table}[t]
\centering
\footnotesize
\begin{tabular}{lccc}
\toprule
code & $m=3$ & $m=4$ & $m=6$ \\
\midrule
surface $d=5$ & $1.015(11)$ & $0.4598(35)$ & $0.456(4)$ \\
HGP(Hamming) & $0.324(4)$ & $0.7(4)$ & $0.3071(30)$ \\
$\mathrm{AT}(3,11)$ & $0.489(5)$ & $5.3(8)$ & $\geq 84$ \\
BB $[[144,12,12]]$ & $1.010(9)$ & $0.58(8)$ & $\geq 12$ \\
BB $[[108,8,10]]$ & $0.87(18)$ & $0.40(8)$ & $\geq 23$ \\
\bottomrule
\end{tabular}
\caption{Effect of the exchange size $m$ at the cluster-exchange probability $p_{\mathrm{cluster}}=0.3$ and the error rate $p=10^{-4}$.
The columns are the relative times $T_{2\%}(m)/T_{2\%}(2)$, the time to accuracy $T_{2\%}$ of \Cref{eq:tta} at the exchange size $m$ divided by that at the exchange size $m=2$, each measured as in \Cref{tab:dmcspeedup} over five runs of $50$ chains.
At the exchange size $m=2$, all five runs reach the $2\%$ target on every code.
Parentheses give bootstrap standard errors over runs, and $\geq$ marks a lower bound where a run did not reach the target within the cap of $L=10^{3}\cdot2^{11}\approx2.0\times10^{6}$ steps.}
\label{tab:clusterk}
\end{table}

\subsection{Exchange size}
\label[appendix]{app:mexch}

The exchange size $m$ is the number of errors that one cluster exchange relocates.
In \Cref{tab:clusterk}, we report the time to accuracy $T_{2\%}$ of \Cref{eq:tta} at the exchange sizes $m=3,4,6$ relative to that at the exchange size $m=2$, with the cluster-exchange probability $p_{\mathrm{cluster}}=0.3$ and the error rate $p=10^{-4}$.

The exchange size $m=3$ takes between $0.32$ and $1.02$ times as long as the exchange size $m=2$ on every code, so either is a reasonable choice.
Which of the exchange sizes $m=2,3,4$ is fastest depends on the code, and the HGP(Hamming) entry for the size $m=4$ contains one slow run out of five.

A cluster exchange of size $m$ needs a configuration of weight at least $m$ and otherwise leaves the configuration unchanged.
At small error rates, the chain stays near the minimum fault weight $w_{\min}$, and at the error rate $p=10^{-4}$ the mean fault weight on the five codes lies between $2.0$ and $3.1$.
On these low-weight configurations, a cluster exchange of size $m=6$ therefore acts as a rejection.
The chain then effectively moves only by the locally balanced flips.
Accordingly, on surface $d=5$ and HGP(Hamming), the two codes without the trapping, the exchange size $m=6$ is faster than the exchange size $m=2$.
On each of the broad-spread codes, $\mathrm{AT}(3,11)$ and the two BB codes, at least one run fails to reach the target, so we keep the exchange size $m=2$ in the main text.

\subsection{Cluster-exchange probability}
\label[appendix]{app:pcluster}

The last free parameter is the probability $p_{\mathrm{cluster}}$ with which \zh{} proposes a cluster exchange instead of a locally balanced flip.
At the exchange size $m=2$ and the error rate $p=10^{-4}$, we sweep the cluster-exchange probability over the values $p_{\mathrm{cluster}}\in\{0,0.1,0.2,0.3,0.5,0.7,0.9\}$ on surface $d=5$, HGP(Hamming), $\mathrm{AT}(3,11)$, and the BB codes $[[144,12,12]]$ and $[[108,8,10]]$ (\Cref{fig:pcluster}).
We obtain the time to accuracy as in \Cref{sec:bench2,tab:dmcspeedup}, over five runs of $50$ chains with the chain length capped at $L=10^{3}\cdot2^{11}\approx2.0\times10^{6}$ steps, and an entry is a lower bound if any run fails to reach the target.

On the two codes without the trapping, surface $d=5$ and HGP(Hamming), \zk{} alone (cluster-exchange probability $p_{\mathrm{cluster}}=0$) is fastest and the time to accuracy $T_{2\%}$ rises monotonically with the cluster-exchange probability.
On these codes, the cluster exchange is only an overhead.

On the broad-spread codes, \zk{} alone did not reach the $2\%$ target within the cap, on $\mathrm{AT}(3,11)$ or either BB code, so a nonzero cluster-exchange probability is needed.
On $\mathrm{AT}(3,11)$, the time to accuracy is smallest at the probability $0.3$ and grows toward both ends of the sweep.
Both BB codes have a broad plateau of nearly equal time to accuracy around the probability $0.3$, spanning four of the seven probabilities.

We therefore use the cluster-exchange probability $p_{\mathrm{cluster}}=0.3$ in the main text for simplicity, although it is not the fastest setting everywhere.

\bibliography{refs}

\end{document}